\documentclass[pra,aps,twocolumn,amssymb,nofootinbib,%
amsmath]{revtex4}
\usepackage{hyperref}
\usepackage{epsfig}
\begin{document}
\title{Electromagnetic Vacuum of Complex Media: Dipole Emission vs. Light Propagation, Vacuum Energy, and Local Field Factors}
\author{M.~Donaire$^{1,2}$}
\email{mdonaire@fc.up.pt}
\affiliation{$^{1}$Centro de F\'{\i}sica do Porto, Faculdade de Ci\^{e}ncias da Universidade do Porto, Rua do Campo Alegre 687, 4169-007 Porto, Portugal\\}

\affiliation{$^{2}$Departamento de F\'{\i}sica de la Materia Condensada, Universidad Aut\'{o}noma de
Madrid, E-28049 Madrid, Spain.}

\date{15 February 2010}

\begin{abstract}
We offer a unified approach to several phenomena related to the electromagnetic vacuum of a complex medium made of point electric dipoles. To this aim, we apply the linear response theory to the computation of the polarization field propagator and study the spectrum of vacuum fluctuations. The physical distinction among the local density of states which enter the spectra of light propagation, total dipole emission, coherent emission, total vacuum energy and Schwinger-bulk energy is made clear. Analytical expressions for the spectrum of dipole emission and for the vacuum energy are derived.
Their respective relations with the spectrum of external light and with the Schwinger-bulk energy are found. The light spectrum and the Schwinger-bulk energy are determined by the Dyson propagator.
The emission spectrum and the total vacuum energy are determined by the polarization propagator.
An exact relationship of proportionality between both propagators is found in terms of local field factors.
A study of the nature of stimulated emission from a single dipole is carried out. Regarding coherent emission, it contains two components. A direct one which is transferred radiatively and directly from the emitter into the medium and whose spectrum is that of external light. And an indirect one which is radiated by induced dipoles. The induction is mediated by one (and only one) local field factor. Regarding the vacuum energy,
we find that in addition to the Schwinger-bulk energy the vacuum energy of an effective medium
contains local field contributions proportional to the resonant frequency and to the spectral line-width.
\end{abstract}
 \maketitle

\section{Introduction}
Dipole emission and the Casimir effects in dielectric media are long-standing research topics in quantum electrodynamics. They constitute paradigmatic phenomena of the interaction matter-radiation related to the quantum nature of the vacuum. Regarding dipole emission, it is known since the work of Purcell \cite{Purcell} that the spontaneous emission rate of an atom, $\Gamma$,
in a dielectric host medium depends on the interaction of the emitter with the material environment. This is so because the back-reaction of the medium on the atom via the self-polarization field modifies its self-energy.  In the first place the host medium modifies the density of channels into which the atom can radiate --i.e. the Local Density of States (LDOS)-- and hence the value of $\Gamma$. Second, the integration of the associated self-energy gives rise to a shift in the resonant frequency of the atom. Third, the dipole-transition-amplitude gets also modified. As a result, the medium is said to renormalize the polarizability of the emitter.\\
\indent In addition and complementarily, the medium polarizes the EM vacuum since material degrees of freedom couple to the EM fluctuations. As a result, polarization fluctuations arise. In particular, this reflects in a renormalization of the susceptibility of the medium, which in turn determines the vacuum energy.\\
\indent We will postulate two distinguishable EM vacua attending to the existence of two different spectra of fluctuations. These are, a \emph{light vacuum} in which the light from an external source propagates through the medium; and an \emph{emission vacuum} in which the radiation emitted from a  dipole within the medium lives. When the dipole constituents of the medium are equivalent to each other, the spectra of fluctuations in both vacua are related by an exact mathematical expression. It will be shown that the spectrum of the light vacuum is only part of the spectrum of the emission vacuum.\\
\indent In accordance to the above distinction, pairs of homologous quantities ascribe to each vacuum. These are, the light propagation spectrum and the dipole emission spectrum; the bulk-Dyson field and the dipole field; the Schwinger-bulk energy and the total vacuum energy. The relationships between the members of each pair are determined by local field factors. The finding of analytical expressions for these relations is the main achievement of the present work.\\
\indent We will deal with a dielectric medium made of isotropic point dipoles. The distribution of dipoles is statistically homogeneous and isotropic. Ours is a classical Green's function based approach. The system is linear, being characterized by a transference matrix $\bar{T}$. The coupling matter-radiation is weak so that in any multiple-scattering process photons are asymptotically free in between any two scattering events. Therefore, the propagator of the radiation in the medium can be computed in terms of a stochastic Lippmann-Schwinger equation over the ensemble of scatterer configurations. Our approach is compatible with the existence of  internal resonances in single scatterers and geometrical resonances in clusters. However, excitons proper of high order systems in the strong coupling regime \cite{Hopfield} are disregarded.\\
\indent The medium under consideration is stationary with respect to (\emph{w.r.t.}) the emission processes. All the quantities computed are time-independent ensemble-averaged quantities. No other mechanism but radiation is considered in the processes of gain and loss of energy by the dipoles. In particular, collisional effects and the Doppler shift are disregarded. For the latter, the condition $v/c\ll\Gamma/\omega$ is required, $v$ being the typical velocity of the scatterers and $\omega$ the frequency of interest. There is also the weak-coupling condition $\omega\gg\Gamma$. Finally, in order to treat the scatterers as point dipoles we demand that $\omega/c\ll a^{-1}$ and $\xi\gtrsim3a$, $a$ being the typical radius of a dipole and $\xi$ being the minimum distance between them.\\
\indent Our approach is based on the microscopical diagrammatic treatments carried out by Foldy \cite{Foldy}, Lax \cite{Lax}, Frisch \cite{Frisch}, Dzyaloshinskii, Gorkov, Abrikosov and Lifshitz \cite{Dzy,Abriko}, Bullough and Hynne \cite{Bullough,BulloughHynne}. We base on Milonni's work \cite{MilonniAmJ,MilonniPRA,MilonniScripta} to apply our classical Green's function formalism to the computation of quantum quantities.\\
\indent The article is organized as follows. In Section \ref{La2} we derive the fundamental relation between the Dyson propagator and the polarization propagator. The self-polarization propagator determines the radiative corrections which renormalize the single-particle polarizability. This issue is addressed in Section \ref{polarizability} together with the derivation of an expression for the optical theorem in complex media. That expression yields the total dipole emission, which is decomposed into coherent and extinguished components in Section \ref{class}. In Section \ref{Vacuum} the light vacuum and the emission vacuum are related via symmetry transformations. The role of the local field factors in coherent emission and field renormalization is explained. In Section \ref{Casimir} we calculate the vacuum energy of a generic complex medium using the Schwinger variational method and particularize to the case of an effective medium. In Section \ref{disc} we comment on the difference between LDOSes and discuss some misconceptions present in previous works. In Section \ref{T} we offer an alternative derivation of the fundamental relation between Dyson's and the polarization propagators based on the Coupled-Dipole-Method. The most relevant results are summarized in the Conclusions.\\
\indent Regarding notation, we will label three-spatial-component vectors with arrows and three-by-three tensors with overlines. We will denote the Fourier-transform of functionals with $q$-dependent arguments instead of the $r$-dependent arguments of their position-space representation.
\section{The Local Density of States}\label{La2}
In application of the fluctuation-dissipation theorem, the LDOS at a point $\vec{r}$ for photons of frequency $\omega$ and any polarization is given by the imaginary part of the trace (Tr) of the $\omega$-mode of the propagator for virtual photons created and annihilated at $\vec{r}$ \cite{Agerwal,Bloch,Economou,Abriko,Landau},
\begin{equation}
\mathcal{N}(\vec{r},\omega)=-\frac{2\omega}{\pi c^{2}}\Im{\Bigl\{\textrm{Tr}\{\bar{\mathbb{G}}(\vec{r},\vec{r};\omega)\}\Bigr\}}.\label{LDOSF}
\end{equation}
When dipoles are present in the medium, in addition to the in-free-space EM fluctuations, polarization fluctuations arise as a result of the interaction matter-radiation. Consequently, the EM vacuum gets polarized and the dielectric matter acquires additional self-energy. The field whose fluctuations in vacuum are named EM vacuum fluctuations is the \emph{self-polarization field}, also called \emph{radiation reaction field} \cite{MilonniAmJ}.\\
\indent Amongst the polarization fluctuations, polaritons are excitations which satisfy certain dispersion relations and propagate coherently \cite{Polaritons}. That is, they are normal modes and they determine the spectrum of coherent dipole emission. We will show in Section \ref{class} that coherent emission consists of two components. Those are, a \emph{direct} component which is transferred radiatively and directly from an emitter into the medium. And an \emph{indirect} component which corresponds to the coherent emission radiated by the induced dipoles surrounding the emitter.\\
\indent We will denote the total LDOS accessible to the photons of frequency $\omega$ emitted from a dipole at $\vec{r}$ by $\mathcal{N}^{emis.}(\vec{r},\omega)$.  We will show that the aforementioned \emph{direct} coherent emission has the same spectrum as the radiation which propagates through the medium from an external (uncorrelated) source. Thus, we will denote by $\mathcal{N}^{light}(\vec{r},\omega)$ that part of $\mathcal{N}^{emis.}(\vec{r},\omega)$ accessible to monochromatic external light of frequency $\omega$. As long as all the dipoles have the same polarizability, any point $\vec{r}$ is statistically equivalent to any other so that we can drop the $\vec{r}$-dependence on both LDOSes in this case.
\subsection{The propagator of the coherent-Dyson field and $\mathcal{N}^{light}$}\label{light}
\begin{figure}[h]
\includegraphics[height=2.1cm,width=6.4cm,clip]{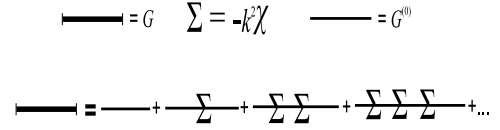}
\caption{Diagrammatic representation
of the Dyson propagator, $\bar{G}$.}\label{FigI03}
\end{figure}

A generic random medium is made of $N+1$ point dipoles or scatterers in a volume $\mathcal{V}$ such that, in the thermodynamical limit $N,\mathcal{V}\rightarrow\infty$, the average numerical density $\rho=\frac{N+1}{\mathcal{V}}$ is finite. Each dipole is labeled by a subscript $i$ which takes integer values in $[0,N]$. Each scatterer configuration is determined by a set of $N+1$ scatterer position vectors $\{\vec{R}_{i}\}$ and a set of $N+1$ renormalized isotropic polarizabilities $\{\bar{\tilde{\alpha}}^{i}\}$ such that $\bar{\tilde{\alpha}}^{i}=\tilde{\alpha}^{i}\bar{\mathbb{I}}$ --see below for the definition of renormalized polarizability. In a statistically homogeneous and isotropic medium all the dipoles have a common polarizability, $\tilde{\alpha}^{i}=\tilde{\alpha}$ $\forall i=0,..,N$, and the position vectors are statistically equivalent as performing ensemble averages. We will stick to a statistically homogenous and isotropic medium unless otherwise specified.\\
\indent $\mathcal{N}^{light}(\omega)$ is given by the Green function of the radiation which propagates from an external source (i.e. ideally located at infinity) through the medium to any point. It is sketched as a series of multiple-scattering diagrams in Fig.\ref{FigI03}. Because both the external source and the end point are spatially uncorrelated \emph{w.r.t.} any of the scattering events in all the diagrams, the corresponding Green's function or propagator is the radiative component of Dyson's, $\bar{G}(\vec{r},\vec{r}';\omega)$ \cite{Dzy,Abriko}. The stochastic Lippmann-Schwinger equation for $\bar{G}(\vec{r},\vec{r}';\omega)$ reads \cite{Frisch},
\begin{widetext}
\begin{equation}\label{bulk}
G_{ij}(\vec{r},\vec{r}';\omega)=G^{(0)}_{ij}(\vec{r}-\vec{r}';\omega)+\int_{\mathcal{V}}\textrm{d}^{3}r^{''}\textrm{d}^{3}r^{'''}
G^{(0)}_{ik}(\vec{r}-\vec{r}^{''};\omega)\Bigl<t_{km}^{\omega}(\vec{r}^{''},\vec{r}^{'''})\Bigr>_{1PI}
G_{mj}(\vec{r}^{'''},\vec{r}';\omega),
\end{equation}
\end{widetext}
where  $\bar{G}^{(0)}(\vec{r}-\vec{r}';\omega)$ is the in-free-space propagator, $k=\omega/c$ and summation over repeated indices is implicit. The functional $\bar{t}^{\omega}(\vec{r}^{''},\vec{r}^{'''})$ denotes the $\bar{t}$-matrix of each specific configuration. It relates to the auxiliary matrix $\bar{\mathrm{t}}^{\omega}(\vec{R}_{i},\vec{R}_{j})$ through
\begin{equation}
\bar{t}^{\omega}(\vec{r}^{''},\vec{r}^{'''})=\sum_{i,j=0}^{N}\bar{\mathrm{t}}^{\omega}(\vec{R}_{i},\vec{R}_{j})\:\delta^{(3)}(\vec{r}^{''}-\vec{R}_{i})\delta^{(3)}(\vec{r}^{'''}-\vec{R}_{j}).
\end{equation}
$\bar{\mathrm{t}}^{\omega}(\vec{R}_{i},\vec{R}_{j})$ propagates photons along all possible multiple-scattering trajectories connecting the dipoles $i$ and $j$ at points $\vec{R}_{i}$ and $\vec{R}_{j}$. It has a cluster expansion, $\bar{\mathrm{t}}^{\omega}(\vec{R}_{i},\vec{R}_{j})=-k^{2}\bar{\tilde{\alpha}}\delta_{ij}+\sum_{n=2}\bar{\mathrm{t}}_{\omega}^{(n)}(\vec{R}_{i},\vec{R}_{j})_{i\neq j}$, where the $n^{th}$ term in the sum contains $n$ factors $-k^{2}\tilde{\alpha}$ and $n-1$ tensors $\bar{G}^{(0)}(\vec{R}_{m}-\vec{R}_{l};\omega)$. As an example,
\begin{eqnarray}
\bar{\mathrm{t}}^{(4)}_{\omega}(\vec{R}_{i},\vec{R}_{j})&=&(-k^{2}\tilde{\alpha})^{4}\sum^{N}_{l,m=0;l\neq m,i;m\neq j}
\bar{G}^{(0)}(\vec{R}_{i}-\vec{R}_{l};\omega)\nonumber\\&\cdot&\bar{G}^{(0)}(\vec{R}_{l}-\vec{R}_{m};\omega)\cdot
\bar{G}^{(0)}(\vec{R}_{m}-\vec{R}_{j};\omega).
\end{eqnarray}
The big brackets in Eq.(\ref{bulk}) stand for the average performed over the ensemble of scatterer-configurations. The subscript 1PI signals the restriction to one-particle-irreducible (1PI) correlations. Thus, the stochastic kernel of Eq.(\ref{bulk}) is the electrical susceptibility function made of the sum of 1PI multiple-scattering processes \cite{Bullough,Frisch},
\begin{equation}
\bar{\chi}^{\omega}(\vec{r}^{''},\vec{r}^{'''})=-k^{-2}\Bigl<\bar{t}^{\omega}(\vec{r}^{''},\vec{r}^{'''})\Bigr>_{1PI}.
\end{equation}
The average process may be performed over both the classical and the quantum degrees of freedom of the configurations. In the simplest case, only spatial correlations among the scatterers are considered \cite{Foldy}.\\
\indent The in-free-space propagator, $\bar{G}^{(0)}(\vec{r}-\vec{r}';\omega)$, is the expectation value of the electric field generated at the point $\vec{r}$ by a classical dipole at $\vec{r}'$ oscillating with frequency $\omega$. Equivalently it is the $\omega-$mode of the Green function of Maxwell's equation in free space,
\begin{equation}\label{Maxwellb}
\Bigl[\frac{\omega^{2}}{c^{2}}\bar{\mathbb{I}}-\vec{\nabla}\times\vec{\nabla}\times\Bigr]\bar{G}^{(0)}(\vec{r}-\vec{r}';\omega)
=\delta^{(3)}(\vec{r}-\vec{r}')\bar{\mathbb{I}}.
\end{equation}
$\bar{G}^{(0)}(\vec{r}-\vec{r}';\omega)$ has also a quantum-mechanical interpretation. It is the scattering amplitude, computed at second-order of perturbation theory, of the interaction between two point dipoles of unit dipole moment placed at $\vec{r}$ and $\vec{r}'$ respectively --see eg. \cite{Andrews}.
$\bar{G}^{(0)}(\vec{r};\omega)$ consists of an electrostatic (Coulombian) dipole field propagator,
\begin{equation}
\bar{G}_{stat.}^{(0)}(r;\omega)=\Bigl[\frac{1}{k^{2}}
\vec{\nabla}\otimes\vec{\nabla}\Bigr]\Bigl(\frac{-1}{4\pi\:r}\Bigr),
\end{equation}
and a radiation field propagator,
\begin{equation}
\bar{G}_{rad.}^{(0)}(r;\omega)=\frac{e^{i\:kr}}{-4\pi
r}\bar{\mathbb{I}}+\bigl[\frac{1}{k^{2}}\vec{\nabla}\otimes\vec{\nabla}\bigr]\frac{e^{i\:kr}-1}{-4\pi r}.
\end{equation}
In the reciprocal space and for an isotropic medium, any tensor can be decomposed into longitudinal and transverse components with respect to the propagation direction, $\vec{q}$.
In free space,
\begin{equation}
\bar{G}^{(0)}(\vec{q};\omega)=G_{\perp}^{(0)}(q;\omega)(\bar{\mathbb{I}}-\hat{q}\otimes\hat{q})+
G_{\parallel}^{(0)}(q;\omega)\:\hat{q}\otimes\hat{q},\nonumber
\end{equation}
where $\hat{q}$ is the unitary vector parallel to $\vec{q}$ and
\begin{equation}\label{losfreek}
G_{\perp}^{(0)}(q;\omega)=\frac{1}{k^{2}-q^{2}},\quad G_{\parallel}^{(0)}(q;\omega)=\frac{1}{k^{2}}.
\end{equation}
While the radiative component is fully transverse, the electrostatic one is fully longitudinal.\\
\indent For a given medium, $\bar{G}(\vec{r},\vec{r}';\omega)$ is the expectation value of the coherent-Dyson field or bulk field, $\vec{E}^{\omega}_{D}(\vec{r})$, generated at $\vec{r}$ by a source at $\vec{r}'$, being both points uncorrelated \emph{w.r.t.} any of the scatterers in the medium. In particular, when $\vec{r}'$ is located at infinity, it is the transverse component \emph{w.r.t.} the propagation, $\bar{G}_{\perp}(\vec{r},\vec{r}';\omega)$, that propagates coherent external light \cite{Bullough}. $\bar{G}$ is obtained averaging over the ensemble of scatterer configurations all possible multiple-scattering propagation trajectories which start at $\vec{r}'$ and end at $\vec{r}$. $\bar{G}(\vec{r},\vec{r}';\omega)$ is said to be the analog of $\bar{G}^{(0)}(\vec{r}-\vec{r}';\omega)$ in a complex medium in the sense that it satisfies an ensemble-averaged Maxwell equation analogous to that in Eq.(\ref{Maxwellb}),
\begin{equation}\label{Maxwellstoch}
\Bigl<\Bigl[\frac{\omega^{2}}{c^{2}}e^{\omega}(\vec{r})\bar{\mathbb{I}}-\vec{\nabla}\times\vec{\nabla}\times\Bigr]\bar{g}(\vec{r},\vec{r}';\omega)\Bigr>
=\delta^{(3)}(\vec{r}-\vec{r}')\bar{\mathbb{I}}.
\end{equation}
In the above expression
$e^{\omega}(\vec{r})=1+\sum^{N}_{i=0}\tilde{\alpha}\delta^{(3)}(\vec{r}-\vec{R}_{i})$ and $\bar{g}(\vec{r},\vec{r}';\omega)$ are respectively the dielectric tensor and the Green function of each specific scatterer-configuration. The integral version of Eq.(\ref{Maxwellstoch}) is that of
Eq.(\ref{bulk}) for $\bar{G}(\vec{r},\vec{r}';\omega)\equiv\Bigl<\bar{g}(\vec{r},\vec{r}';\omega)\Bigr>$.\\
\indent In a statistically homogeneous medium $\bar{G}(\vec{r},\vec{r}';\omega)$ and $\bar{\chi}^{\omega}(\vec{r},\vec{r}')$ are functions of $\vec{r}-\vec{r}'$ for any pair of points. In the Fourier space, isotropy allows to split the Dyson
equation for $\bar{G}(q)$ in two uncoupled and mutually orthogonal algebraic equations,
\begin{equation}
G_{\perp,\parallel}(q)=G_{\perp,\parallel}^{(0)}(q)\:-\:k^{2}\:G_{\perp,\parallel}^{(0)}(q)\:\chi_{\perp,\parallel}(q)\:G_{\perp,\parallel}(q).\label{DysonI}
\end{equation}
In the above equation and in the following we will omit for brevity the explicit dependence of the functionals on $\omega$.
The same as for the $\bar{t}$-matrix, $\chi_{\perp,\parallel}(q)$  adjust to cluster expansions of the form,
\begin{equation}\label{laXenAes}
\chi_{\perp,\parallel}(q)=\sum_{n=1}^{\infty}X^{(n)}_{\perp,\parallel}(q)\rho^{n}
\tilde{\alpha}^{n}.
\end{equation}
The functions
$X^{(n)}_{\perp,\parallel}(q)$  incorporate the spatial dispersion due to the 1PI spatial correlations within clusters of $n$ scatterers. In addition, they account for the intermediate
multiple-recurrent-scattering processes among them. In field theory terminology, $\bar{\chi}$ is proportional to the \emph{self-energy tensor}, $\Sigma_{\perp,\parallel}(q)=-k^{2}\chi_{\perp,\parallel}(q)$. Alternatively, Eq.(\ref{DysonI}) can be written in terms of the ensemble-averaged $\bar{T}$-matrix,
\begin{equation}
G_{\perp,\parallel}(q)=G_{\perp,\parallel}^{(0)}(q)\:+\:G_{\perp,\parallel}^{(0)}(q)\:T_{\perp,\parallel}(q)\:
G^{(0)}_{\perp,\parallel}(q),\label{Dysonalter}
\end{equation}
where, in application of isotropy and homogeneity, $T_{\perp,\parallel}$ are defined through
\begin{eqnarray}
\Bigl<\bar{t}(\vec{q},\vec{q}')\Bigr>&=&
(2\pi)^{3}\delta^{(3)}(\vec{q}-\vec{q}')\bar{T}(\vec{q}),\nonumber\\
\bar{T}(\vec{q})&=&T_{\perp}(q)(\bar{\mathbb{I}}-\hat{q}\otimes\hat{q})+
T_{\parallel}(q)\:\hat{q}\otimes\hat{q}.\nonumber
\end{eqnarray}
The solutions to Eq.(\ref{DysonI}) are the Dyson propagator components,
\begin{eqnarray}\label{effectivG}
G_{\perp}(q)&=&\frac{1}{k^{2}[1+\chi_{\perp}(q)]-q^{2}},\nonumber\\
G_{\parallel}(q)&=&\frac{1}{k^{2}[1+\chi_{\parallel}(q)]}.
\end{eqnarray}
In terms of free propagators and self-energy tensors, $\bar{G}$ can be depicted perturbatively as in Fig.\ref{FigI03}.
Normal modes are given by the poles of $G_{\perp,\parallel}(q)$. They satisfy the dispersion relations,
\begin{eqnarray}
k^{2}\epsilon_{\perp}(q)-q^{2}|_{q=k^{nor.}_{\perp}}=0,\label{qt}\\
\epsilon_{\parallel}(q)|_{q=k^{nor.}_{\parallel}}=0,\label{ql}
\end{eqnarray}
with $\epsilon_{\perp,\parallel}(q)=1+\chi_{\perp,\parallel}(q)$ being the dielectric functions for transverse and longitudinal modes respectively. Longitudinal modes need of coupling to matter to propagate while external light excites only transverse modes \cite{Bullough}. Thus, we have
\begin{equation}
\mathcal{N}^{light}(\omega)=-\frac{4\omega}{\pi c^{2}}\Im{\Bigl\{\int\frac{\textrm{d}^{3}q}{(2\pi)^{3}}G_{\perp}(q;\omega)\Bigr\}}.\label{LDOSsl}
\end{equation}

\subsection{The propagator of the polarization field and $\mathcal{N}^{emis.}$}\label{2B}
\begin{figure}[h]
\includegraphics[height=4.9cm,width=7.8cm,clip]{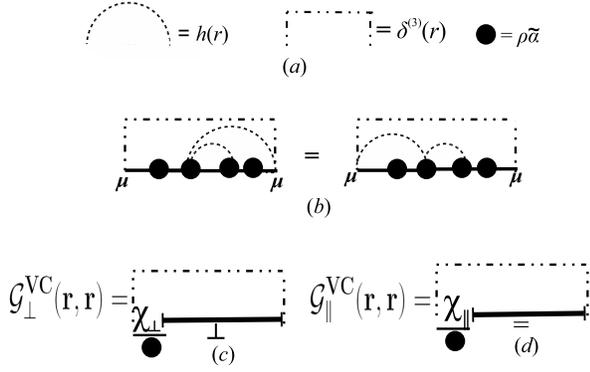}
\caption{($a$) Feynman rules. Only two-point irreducible correlation functions, $h(r)$, have been used for the sake of simplicity. ($b$) Diagrammatic representation of the
equivalence between multiple-scattering processes amounting to
$\bar{\mathcal{G}}(\vec{r},\vec{r})$. Reciprocity applies. $\mu$ represents the virtual presence of the emitter. ($c$),($d$) Diagrammatic representations of the integration in $q$ of Eq.(\ref{LDOSIper}) and Eq.(\ref{LDOSIparal}) respectively.}\label{FigI04}
\end{figure}

The Green function which determines $\mathcal{N}^{emis.}(\omega)$ through Eq.(\ref{LDOSF}) is that of the polarization field, $\bar{\mathcal{G}}$. For a given medium, $\bar{\mathcal{G}}(\vec{r},\vec{r}';\omega)$  is the expectation value of the electric field generated at the point $\vec{r}$ by a classical dipole oscillating at frequency $\omega$ at a point $\vec{r}'$ within the medium. It is in this sense that $\bar{\mathcal{G}}(\vec{r},\vec{r}';\omega)$ is the analog of $\bar{G}^{(0)}(\vec{r}-\vec{r}';\omega)$ in a complex medium. Because, by assumption, our medium is made of indistinguishable point scatterers we choose the dipole emitter to be the $0^{th}$ scatterer out of the $N+1$. Thus, the ensemble-averaged Maxwell equation reads here,
\begin{equation}\label{Polstoch}
\Bigl<\Bigl[\frac{\omega^{2}}{c^{2}}e^{\omega}(\vec{r})\bar{\mathbb{I}}-\vec{\nabla}\times\vec{\nabla}\times\Bigr]\bar{\mathfrak{g}}(\vec{r},\vec{r}';\omega)\Bigr>
\Bigr|_{\vec{r}'=\vec{R}_{0}}
=\delta^{(3)}(\vec{r}-\vec{R}_{0})\bar{\mathbb{I}}.
\end{equation}
The restriction $|_{\vec{r}'=\vec{R}_{0}}$  signals the fact that the ensemble-averaged is performed keeping $\vec{r}'$ fixed at the position vector of the $0^{th}$ scatterer. This simple but crucial fact makes Eq.(\ref{Polstoch}) different to Eq.(\ref{Maxwellstoch}). The resultant integral Lippmann-Schwinger stochastic equation for $\bar{\mathcal{G}}(\vec{r},\vec{r}';\omega)=\Bigl<\bar{\mathfrak{g}}(\vec{r},\vec{r}';\omega)\Bigr>$ is,
\begin{widetext}
\begin{equation}\label{emit}
\mathcal{G}_{ij}(\vec{r},\vec{r}';\omega)=G^{(0)}_{ij}(\vec{r}-\vec{r}';\omega)+\int_{\mathcal{V}}\textrm{d}^{3}r^{''}\textrm{d}^{3}r^{'''}
G^{(0)}_{ik}(\vec{r}-\vec{r}^{''};\omega)\Bigl<t^{\omega}_{km}(\vec{r}^{''}-\vec{r}',\vec{r}^{'''}-\vec{r}')
\Bigr>_{1PI}\Bigr|_{\vec{r}'=\vec{R}_{0}}\mathcal{G}_{mj}(\vec{r}^{'''},\vec{r}';\omega).
\end{equation}
\end{widetext}
The expression
\begin{equation}\label{skernel}
\Bigl<t^{\omega}_{km}(\vec{r}^{''}-\vec{r}',\vec{r}^{'''}-\vec{r}')
\Bigr>_{1PI}\Bigr|_{\vec{r}'=\vec{R}_{0}}
\end{equation}
is a symbolic manner of writing the stochastic kernel. The explicit dependence of $\bar{t}^{\omega}$ on $\vec{r}'$ and the restriction $|_{\vec{r}'=\vec{R}_{0}}$ denote that while performing the average over all the 1PI processes, $\vec{r}'$ in $\bar{\mathcal{G}}(\vec{r}^{'''},\vec{r}';\omega)$ is kept fixed at the center of the $0^{th}$ dipole. This is a consequence of the fact that the stochastic operator on the \emph{l.h.s.} of Eq.(\ref{Polstoch}) is correlated to the source term on the \emph{r.h.s} of that equation. In Eq.(\ref{emit}) the dipole at $\vec{r}'$ acts as a source and hence it is a dipole emitter. In general, the presence of an emitter would break statistical homogeneity in a random medium and $\bar{\mathcal{G}}$ would depend explicitly on the emitter location, $\vec{R}_{0}$. However in a statistically homogeneous medium, since all the position vectors are  equivalent as performing ensemble averages, we can drop the explicit dependence on $\vec{R}_{0}$ in favor of a generic argument $\vec{r}'$ as in Eq.(\ref{emit}). Note that if we replace $\vec{r}'$ by $\vec{r}$ in the stochastic kernel and fix $\vec{r}$ at a dipole, that dipole would be a particle being polarized instead.\\
\indent Our interest is in the self-polarization propagator that enters $\mathcal{N}^{emis.}(\omega)$. It is computed out of Eq.(\ref{emit}) by setting  $|\vec{r}'-\vec{r}|<a$. This way, the dipole emitter is at the same time the dipole being polarized. In turn, the distinction between $\vec{r}$ and $\vec{r}'$ is just formal as it cannot be resolved for point dipoles. We will set  $\vec{r}=\vec{r}'$ without loss of generality. $\bar{\mathcal{G}}(\vec{r},\vec{r};\omega)$  carries the radiative corrections which enter the renormalized polarizability of the emitter, $\tilde{\alpha}$. Because both the fluctuations of the classical self-polarization field and those of the quantum vacuum field are related by the fluctuation-dissipation theorem --see eg.\cite{MilonniAmJ,MilonniPRA}, both classical and quantum corrections can be treated within the same formalism. Clarifications on this point will be provided where necessary.\\
\indent The explicit dependence of Eq.(\ref{skernel}) on the emitter position in Eq.(\ref{emit}) makes that any scattering process of the virtual photons in their way from and towards the emitter be correlated \emph{w.r.t.} the emitter position. When the emitter is in all equivalent to the rest of host scatterers, such a correlation is the same as that among the host scatterers themselves. In the simplest case the only correlation is that of an exclusion volume. It is in this sense that the homogeneity of the medium is only virtually broken and the exclusion volume around the emitter is referred to as \emph{virtual cavity}. Should the emitter be distinguishable \emph{w.r.t.} the other scatterers, that correlation would differ. Homogeneity would be actually broken and the cavity would not be
virtual but a \emph{real cavity}. The breaking of homogeneity, either virtually or actually, makes the computation of $\bar{\mathcal{G}}(\vec{r},\vec{r}';\omega)$ different to that of the ordinary Dyson's propagator and the stochastic kernel of Eq.(\ref{emit}) different to the susceptibility tensor. We will refer generically to the scenario in which the emitter is in all equivalent to the host dipoles as \emph{stricto sensu} (s.s.) virtual cavity (VC) scenario. It is for this scenario that there exists a close relation between $\bar{G}(\vec{r},\vec{r};\omega)$ and $\bar{\mathcal{G}}(\vec{r},\vec{r};\omega)$ which we proceed to investigate.\\
\indent In the following, we will work in Fourier space and write $\bar{\mathcal{G}}(\vec{r},\vec{r};\omega)$ as
\begin{eqnarray}
\bar{\mathcal{G}}(\vec{r},\vec{r};\omega)&=&\frac{1}{3}\Bigl[\int\frac{\textrm{d}^{3}q}{(2\pi)^{3}}2\mathcal{G}_{\perp}(q;\omega)
+\int\frac{\textrm{d}^{3}q}{(2\pi)^{3}}\mathcal{G}_{\parallel}(q;\omega)\Bigr]\bar{\mathbb{I}}\nonumber\\&\equiv&
\frac{1}{3}\Bigl[[2\varphi_{\perp}^{\omega(0)}+\varphi_{\parallel}^{\omega(0)}]+2\varphi^{\omega sc}_{\perp}+\varphi^{\omega sc}_{\parallel}\Bigr]\bar{\mathbb{I}},\label{gamy}
\end{eqnarray}
where $\varphi_{\perp,\parallel}^{\omega(0)}$ are the in-free-space values which contain real divergences to be regularized. $\varphi^{\omega sc}_{\perp,\parallel}$ are the divergenceless scattering pieces. Physically, the $\varphi$-factors account for the dipole self-energy associated to self-polarization photons. The numerical prefactor 2 in front of $\varphi_{\perp}$ stands for the two transverse polarization modes. The $\omega$-dependence will be omitted hereafter unless necessary.
We first observe that $\bar{\mathcal{G}}(\vec{r},\vec{r})$ is made of 1PI diagrams in which the end points coincide. Those diagrams amount to the so-called recurrent scattering. Let us draw them as linear processes as in Fig.\ref{FigI04}. The second observation is that, by reciprocity,  every intermediate scattering event in $\bar{\mathcal{G}}(\vec{r},\vec{r})$ which is correlated to the fixed dipole at $\vec{r}$ will be correlated either at the near end or at the rare end of each diagram, indistinguishably. Taking advantage of this feature in every multiple-scattering diagram like that in Fig.\ref{FigI04}($b$), we can attribute all the irreducible correlations of the intermediate scattering events to the emitter on the
left. By proceeding so, we end up with an effective separation of all those pieces irreducibly correlated to the emitter on the left completely untangled from those non-1PI pieces on the right. The sum of the 1PI pieces on the left amounts to  $\bar{\chi}/\rho\tilde{\alpha}$, where the factor $1/\rho\tilde{\alpha}$ stands for the removal of the first random scatter which enters the diagrams of $\bar{\chi}$ in favor of the fixed emitter location.
The sum of the non-1PI pieces on the right amounts to the bulk propagator $\bar{G}$. As a result, we end up with the formulae --Figs.\ref{FigI04}($c,d$),
\begin{eqnarray}
\mathcal{G}^{VC}_{\perp}(q)&=&\frac{\chi_{\perp}(q)}{\rho\tilde{\alpha}}G_{\perp}(q)=
\frac{\chi_{\perp}(q)/(\rho\tilde{\alpha})}{k^{2}[1+\chi_{\perp}(q)]-q^{2}},\label{LDOSIper}\\
\mathcal{G}^{VC}_{\parallel}(q)&=&\frac{\chi_{\parallel}(q)}{\rho\tilde{\alpha}}\:G_{\parallel}(q)
=\frac{1}{\rho\tilde{\alpha}}\:\frac{\chi_{\parallel}(q)}{k^{2}[1+\chi_{\parallel}(q)]}.\label{LDOSIparal}
\end{eqnarray}
The above expressions contain both $G_{\perp}^{(0)}$ and $G_{\parallel}^{(0)}$ which carry divergences. The scattering terms,
\begin{eqnarray}
\varphi^{VC}_{\perp}&=&
\int\frac{\textrm{d}^{3}q}{(2\pi)^{3}}\Bigl[\frac{\chi_{\perp}(q)/(\rho\tilde{\alpha})}{k^{2}
[1+\chi_{\perp}(q)]-q^{2}}-G_{\perp}^{(0)}(q)\Bigr],
\label{LDOSIperg}\\
\varphi^{VC}_{\parallel}&=&\int\frac{\textrm{d}^{3}q}{(2\pi)^{3}}\:\Bigl[\frac{1}{\rho\tilde{\alpha}}
\frac{\chi_{\parallel}(q)}{k^{2}[1+\chi_{\parallel}(q)]}-G_{\parallel}^{(0)}(q)\Bigr],\label{LDOSIparalg}
\end{eqnarray}
are however fully convergent.\\
\indent Alternatively, we can write $\bar{\mathcal{G}}^{VC}(q)$ in other forms making use of  Dyson's equation. Using Eq.(\ref{DysonI}) we get,
\begin{equation}\label{infunctionofG}
\mathcal{G}^{VC}_{\perp,\parallel}(q)=\frac{1}{k^{2}\rho\tilde{\alpha}}
\Bigl[1-\frac{G_{\perp,\parallel}}{G_{\perp,\parallel}^{(0)}}\Bigr].
\end{equation}
In function of the $\bar{T}$-matrix of Eq.(\ref{Dysonalter}) it reads,
\begin{equation}\label{infunctionofT}
\mathcal{G}^{VC}_{\perp,\parallel}(q)=\frac{-1}{k^{2}\rho\tilde{\alpha}}\:
T_{\perp,\parallel}(q)\:G^{(0)}_{\perp,\parallel}(q).
\end{equation}
The reader can find in Section \ref{T} an alternative proof of Eq.(\ref{infunctionofT}) using the formalism of the Coupled-Dipole-Method.\\
\indent Finally, we write the Lippmann-Schwinger equation of Eq.(\ref{emit}) for the polarization field propagator in Fourier space,
\begin{equation}
\mathcal{G}^{VC}_{\perp,\parallel}(q)=G_{\perp,\parallel}^{(0)}(q)+G_{\perp,\parallel}^{(0)}(q)\:\Xi^{VC}_{\perp,\parallel}(q)\:\mathcal{G}^{VC}_{\perp,\parallel}(q),\label{DysonImathG}
\end{equation}
where
\begin{equation}\label{Xi}
\Xi^{VC}_{\perp,\parallel}(q)=\frac{-\rho\tilde{\alpha}}{\chi_{\perp,\parallel}\:G^{(0)}_{\perp,\parallel}}
\Bigl[1\:-\:\frac{\chi_{\perp,\parallel}}{\rho\tilde{\alpha}}\:+\:k^{2}\chi_{\perp,\parallel}\:G^{(0)}_{\perp,\parallel}\Bigr](q)
\end{equation}
is the stochastic kernel of Eq.(\ref{skernel}) expressed as a function of $\bar{\chi}$, $\bar{G}^{(0)}$ and $\rho\tilde{\alpha}$.\\
\indent In terms of the $\varphi$-factors defined in Eq.(\ref{gamy}), the LDOS of emission reads,
\begin{equation}
\mathcal{N}^{emis.}(\omega)=-\frac{2\omega}{\pi c^{2}}\Im{\Bigl\{2\varphi_{\perp}^{\omega(0)}
+2\varphi^{\omega sc}_{\perp}+\varphi^{\omega sc}_{\parallel}\Bigr\}}.\label{LDOSsp}
\end{equation}
Note that, differently to the case of $\mathcal{N}^{light}(\omega)$, even for the case that longitudinal normal modes do not exist, both transverse and longitudinal modes contribute generally to $\mathcal{N}^{emis.}(\omega)$ as it does not solely include coherent modes. This being so regardless of the absorptive properties of the medium.

\section{Renormalization of the single-particle polarizability}\label{polarizability}
The propagator $\bar{\mathcal{G}}(\vec{r},\vec{r};\omega)$ carries the radiative corrections which renormalize the so-called bare-electrostatic polarizability $\alpha_{0}$ of a given dipole. $\alpha_{0}$ can take different forms depending on the nature of the dipole. We refer to \cite{RMPdeVries} for a comprehensive summary. In Appendix \ref{append} we  adopt a semiclassical model in which dipoles are spherical nanoparticles of permittivity $\epsilon_{e}(\omega)$ and radius $a\ll c/\omega$ with $\alpha_{0}(\omega)=4\pi a^{3}\frac{\epsilon_{e}(\omega)-1}{\epsilon_{e}(\omega)+2}$. In Appendix \ref{appendy} we adopt a two-level atom model in which the atom is in a spherically symmetric quantum state.\\
\indent Let us consider first the classical model and compute the off-resonant renormalized polarizability of a classical dipole. Let us suppose that the dipole is stimulated by a stationary external field which oscillates in time with  frequency $\omega$, $\vec{E}_{0}$. The processes of absorbtion and emission of radiation by the emitter become stationary after a relaxation time of the order of $\Gamma^{-1}$.
In function of the $\varphi$-factors, the average power emitted and absorbed by the stimulated dipole reads --see Appendix \ref{append},
\begin{eqnarray}
W^{tot}_{\omega}&=&
\frac{\omega\epsilon_{0}}{2}\Im{}\Bigl\{\frac{\alpha_{0}}{1+\frac{1}{3}k^{2}\alpha_{0}[2\varphi^{(0)}_{\perp}+2\varphi^{sc}_{\perp}+
\varphi^{sc}_{\parallel}]}
\Bigr\}|\vec{E}_{0}|^{2}\nonumber\\&=&\frac{-\omega^{3}\epsilon_{0}}{6c^{2}}\Bigl[\frac{|\alpha_{0}|^{2}
\Im{\{2\varphi_{\perp}^{(0)}+2\varphi^{sc}_{\perp}
+\varphi^{sc}_{\parallel}\}}}
{|1+\frac{1}{3}k^{2}\alpha_{0}[2\varphi^{(0)}_{\perp}+2\varphi^{sc}_{\perp}+\varphi^{sc}_{\parallel}]|^{2}}\label{la2}\\
&-&\frac{3}{k^{2}}
\frac{\Im{\{\alpha_{0}\}}}{|1+\frac{1}{3}k^{2}\alpha_{0}[2\varphi^{(0)}_{\perp}+2\varphi^{sc}_{\perp}+\varphi^{sc}_{\parallel}]|^{2}}\Bigr]
|\vec{E}_{0}|^{2}.\label{la3}
\end{eqnarray}
The explicit dependence of the $\varphi$-factors on $k=\omega/c$ has been omitted.
The term in Eq.(\ref{la3}) corresponds to the power
 absorbed within the emitter while that of Eq.(\ref{la2}) corresponds to the power transferred to the medium.\\
\indent We can write $W^{tot}_{\omega}$ in terms of a renormalized polarizability $\tilde{\alpha}$ as $W^{tot}_{\omega}=\frac{\omega\epsilon_{0}}{2}\Im{\{\tilde{\alpha}\vec{E}_{0}\cdot\vec{E}_{0}^{*}\}}=
\frac{\omega\epsilon_{0}}{2}|\vec{E}_{0}|^{2}\Im{\{\tilde{\alpha}\}}$, with
\begin{equation}\label{alpha1}
\tilde{\alpha}(k)=\frac{\alpha_{0}}{1+\frac{2}{3}k^{2}\alpha_{0}\varphi^{(0)}_{\perp}+\frac{1}{3}k^{2}\alpha_{0}
[2\varphi^{sc}_{\perp}+\varphi^{sc}_{\parallel}]}.
\end{equation}
Because the computation has been assumed for frequencies far from resonances, it is possible to regularize the real divergence of $\varphi^{(0)}_{\perp}$ by setting it to zero \cite{RMPdeVries} so that $2\varphi^{(0)}_{\perp}=-i\frac{k}{2\pi}$ in the above equation.
Finally, writing Eqs.(\ref{la2},\ref{la3}) in the form,
\begin{eqnarray}\label{laWshort}
W^{tot}_{\omega}=
\frac{-\omega^{3}\epsilon_{0}}{6c^{2}}|\tilde{\alpha}\vec{E}_{0}|^{2}\Bigl[\Im{\{2\varphi^{(0)}_{\perp}+2\varphi^{sc}_{\perp}
+\varphi^{sc}_{\parallel}\}}-\frac{3}{k^{2}}\frac{\Im{\{\alpha_{0}\}}}{|\alpha_{0}|^{2}}\Bigr],
\end{eqnarray}
we obtain an expression for the optical theorem in complex media.\\
\indent Next, let us consider the polarizability of a quantum two-level atom. In good approximation, we can adjust $\tilde{\alpha}$ to a renormalized Lorentzian polarizability \cite{PRLdeVries},
\begin{equation}\label{Lorentzian}
\tilde{\alpha}=\frac{1}{3}\tilde{\alpha}_{0}k_{res}^{2}[k_{res}^{2}-k^{2}-i\Gamma k^{3}/(c k_{res}^{2})]^{-1}.
\end{equation}
This way, $\tilde{\alpha}$ gets parametrized in terms of the renormalized values of $\tilde{\alpha}_{0}$, $k_{res}$ and $\Gamma$. Radiative corrections enter in the same fashion as for the classical model above \cite{MilonniAmJ} --see also Appendix \ref{appendy}. Therefore, the parametrization in Eq.(\ref{Lorentzian}) allows us to give a physical meaning to the $\varphi$-factors and to their in-free-space divergent terms. In free space $\alpha_{0}$ relates to the dipole transition matrix element $\mu$ and the bare resonant frequency, $\omega_{0}=c k_{0}$, through $\alpha_{0}=\frac{2|\mu|^{2}}{\epsilon_{0}\hbar c k_{0}}$. Also, the above parametrization yields an automatic regularization of the divergent real part of $2\varphi^{(0)}_{\perp}$, $\Re{\{2\varphi_{\perp}^{(0)}\}}=\frac{-3}{k_{0}^{2}\alpha_{0}}$. This regularization procedure determines the in-free-space Lamb-shift when coupling to bare radiative modes is considered \cite{Wylie}. By consistency,  $\Gamma_{0}=c\alpha_{0}k_{0}^{4}/6\pi=\frac{k_{0}^{3}}{3\pi\epsilon_{0}\hbar}|\mu|^{2}$ is the in-free-space decay rate. Comparison of Eq.(\ref{alpha1}) with Eq.(\ref{Lorentzian}) yields the following expressions for the renormalized parameters in terms of the $\varphi$-factors,
\begin{eqnarray}\label{Gamares}
\Gamma&=&-\frac{c}{3}\tilde{\alpha}_{0}k^{3}\Im{\{2\varphi^{(0)}_{\perp}+2\varphi^{sc}_{\perp}+\varphi^{sc}_{\parallel}\}}|_{k=k_{res}}
\nonumber\\&=&-\Gamma_{0}\frac{2\pi}{k_{0}^{2}}k\Im{\{2\varphi^{(0)}_{\perp}+2\varphi^{sc}_{\perp}+\varphi^{sc}_{\parallel}\}}|_{k=k_{res}},
\end{eqnarray}
where $k_{res}$ is a real non-negative root of the equation
\begin{equation}\label{kres}
(k/k_{0})^{2}-1=\frac{1}{3}\alpha_{0}k^{2}\Re{\{2\varphi^{sc}_{\perp}+\varphi^{sc}_{\parallel}\}}|_{k=k_{res}},
\end{equation}
\begin{equation}\label{alpha0reg}
\textrm{ and }\quad\tilde{\alpha}_{0}=\alpha_{0}(k_{0}/k_{res})^{2}.
\end{equation}
Except for Eq.(\ref{alpha0reg}), the formulae for $k_{res}$ and $\Gamma$ have been already derived using a fully quantum-mechanical (QM) formalism \cite{Wylie,WylieII,WelschI,WelschII}. In \cite{WelschI,WelschII} the same self-consistency requirement between Eq.(\ref{Gamares}) and Eq.(\ref{kres}) was obtained. Note however that only the resonant term of the level shift is accounted for in Eq.(\ref{kres}) \cite{WelschI}. That is, we have implicitly assumed that the level shift due to the van-der-Waals interactions is the same for the two atomic levels \cite{WylieII}. See also Section \ref{Casimir}.\\
\indent The underlying reason why the radiative corrections which yield the spectrum of stimulated emission in Eq.(\ref{laWshort}) enter in the same fashion the spontaneous decay rate in Eq.(\ref{Gamares}) is that the spectrum of the self-polarization field relates to that of the vacuum field by means of the fluctuation-dissipation theorem. While it can be interpreted that only the latter contributes to Eq.(\ref{laWshort}), both of them, and in the same fashion, contribute to Eq.(\ref{Gamares}). Following Milonni's arguments \cite{MilonniAmJ,MilonniScripta}, this explains why the Einstein coefficient for spontaneous emission is twice that for stimulated emission.

\section{On the nature of stimulated emission}\label{class}
Let us consider the emission term of Eq.(\ref{la2}) for a scenario in which only one of the classical dipoles of a statistically homogeneous dielectric is stimulated by an external field. We will refer to the emitter as dipole source as well. The $\varphi$-factors are those for the virtual cavity scenario and we can write,
\begin{equation}
W^{tot}_{w}=W_{o}\int\frac{\textrm{d}^{3}q}{(2\pi)^{3}}
\textrm{Tr}\Bigl\{\Im{\{\frac{\bar{\chi}^{\omega}(\vec{q})}{\rho\tilde{\alpha}}\cdot
\bar{G}(\vec{q};\omega)\}}\Bigr\}\label{GVC},
\end{equation}
where $W_{o}=\frac{-\omega^{3}}{6c^{2}\epsilon_{0}}|\vec{p}_{0}|^{2}$, being $\vec{p}_{0}=\epsilon_{0}\tilde{\alpha}\vec{E}_{0}(\vec{r})$ the dipole moment induced by the monochromatic external field $\vec{E}_{0}(\vec{r})$ of frequency $\omega$ on the emitter at $\vec{r}$.\\
\indent The first obvious decomposition is that between transverse, $W^{tot}_{\perp}$, and longitudinal emission, $W^{tot}_{\parallel}$,
\begin{eqnarray}
W^{tot}_{\perp}&=&W_{o}\int\frac{\textrm{d}^{3}q}{(2\pi)^{3}}\:2\Im{\{\frac{\chi_{\perp}(q)}{\rho\tilde{\alpha}}G_{\perp}(q)\}},\label{eka}\\
W^{tot}_{\parallel}&=&W_{o}\int\frac{\textrm{d}^{3}q}{(2\pi)^{3}}\:\Im{\{\frac{\chi_{\parallel}(q)}{\rho\tilde{\alpha}}G_{\parallel}(q)\}}.\label{eke}
\end{eqnarray}
Hereafter we drop the script $\omega$ bearing in mind that all the quantities depend on the frequency of the external field.\\
\indent Coherent emission is that carried by the coherent-Dyson field itself and by any other averaged-field component \emph{in-phase} with it. That is, the power carried by the total ensemble-averaged field radiated \cite{Lax}. This is the radiation whose modes satisfy the same dispersion relations as the normal modes in the bulk, Eqs.(\ref{qt},\ref{ql}) \cite{Bullough,Hopfield,Fano}.
They are the poles of $G_{\perp}(q)$ and $G_{\parallel}(q)$ in the integrands of Eqs.(\ref{eka},\ref{eke}). The rest of the radiation goes into dispersion and absorbtion. It is termed generically \emph{extinguished} emission \cite{Sentenac}. Therefore, we define,
\begin{eqnarray}
W^{Coh.}_{\perp}&=&W_{o}\int\frac{\textrm{d}^{3}q}{(2\pi)^{3}}\:2\Re{\{\frac{\chi_{\perp}(q)}{\rho\tilde{\alpha}}\}}\Im{\{G_{\perp}(q)\}},
\label{Gammapcoh}\\
W^{ext.}_{\perp}&=&W_{o}\int\frac{\textrm{d}^{3}q}{(2\pi)^{3}}\:2\Im{\{\frac{\chi_{\perp}(q)}{\rho\tilde{\alpha}}\}}\Re{\{G_{\perp}(q)\}},
\label{Gammapincoh}\\
W^{Coh.}_{\parallel}&=&W_{o}\int\frac{\textrm{d}^{3}q}{(2\pi)^{3}}\:\Re{\{\frac{\chi_{\parallel}(q)}{\rho\tilde{\alpha}}\}}\Im{\{G_{\parallel}(q)\}},
\label{Gammalcoh}\\
W^{ext.}_{\parallel}&=&W_{o}\int\frac{\textrm{d}^{3}q}{(2\pi)^{3}}\:\Im{\{\frac{\chi_{\parallel}(q)}{\rho\tilde{\alpha}}\}}\Re{\{G_{\parallel}(q)\}}.
\label{Gammalincoh}
\end{eqnarray}
The definitions of formulae Eqs.(\ref{Gammapcoh},\ref{Gammalcoh}) as transverse/longitudinal coherent emission are inspired by the spectrum of the on-shell modes of field theories \cite{Abriko,Peskin}. Hence, the sourceless counterpart of $W^{Coh.}_{\omega}=W^{Coh.}_{\perp}+W^{Coh.}_{\parallel}$ is the spectrum of polaritons. In the following, we examine $W^{Coh.}_{\omega}$ in the framework of Classical Optics to show that it corresponds indeed to the ordinary classical interpretation of coherent radiation. Let us write $W^{Coh.}_{\omega}$ as
\begin{eqnarray}
W^{Coh.}_{\omega}&=&\frac{-\omega^{3}}{6c^{2}\epsilon_{0}}|\vec{p}_{0}|^{2}\label{Wcoher}\\&\times&\int \textrm{d}^{3}r'\textrm{Tr}\Bigl\{\Re{\{\bar{\chi}(\vec{r}-\vec{r}')/\rho\tilde{\alpha}\}}\cdot
\Im{\{\bar{G}(\vec{r}',\vec{r})\}}\Bigr\}\nonumber,
\end{eqnarray}
where both $\bar{\chi}$ and $\bar{G}$ are written in the position-space representation for convenience. Considering the fields classically, the fluctuation-dissipation relation reads \cite{Economou,Abriko,Landau},
\begin{equation}\label{Disiclass}
\Im{\{\bar{G}(\vec{r}',\vec{r};\omega)\}}=-\frac{\pi\epsilon_{0}}{\hbar k^{2}}
\langle\vec{E}^{\omega}_{D}(\vec{r}')\otimes\vec{E}^{\omega*}_{D}(\vec{r})\rangle,
\end{equation}
where $\vec{E}^{\omega}_{D}(\vec{r})$ is the $\omega$-mode of the coherent-Dyson field\footnote{Throughout this paper the electric field vectors with script $\omega$ are frequency modes of the field, not to be confused with the monochromatic external electric field of frequency $\omega$ used in Sections \ref{polarizability},\ref{T}.} and the script $D$ stands both for Dyson and for \emph{direct} emission for the reasons explained below. Using Eq.(\ref{Disiclass}) and writing $\Re{\{\bar{\chi}(\vec{r}-\vec{r}')/\rho\tilde{\alpha}\}}=[\Re{\{\bar{\chi}(\vec{r}-\vec{r}')/\rho\tilde{\alpha}\}}
-\delta^{(3)}(\vec{r}-\vec{r}')\bar{\mathbb{I}}]\:+\:\delta^{(3)}(\vec{r}-\vec{r}')\bar{\mathbb{I}}$ in Eq.(\ref{Wcoher}), we separate explicitly the field radiated directly by the source dipole from that which is  emitted by the induced  surrounding dipoles,
\begin{eqnarray}
W^{Coh.}_{\omega}&=&\frac{\pi\omega}{6\hbar}|\vec{p}_{0}|^{2}\langle\vec{E}^{\omega}_{D}(\vec{r})\cdot\vec{E}^{\omega*}_{D}(\vec{r})\rangle\nonumber\\
&+&\frac{\pi\omega}{6\hbar}|\vec{p}_{0}|^{2}\textrm{Tr}\Bigl\{\int\textrm{d}^{3}r'\Bigl\langle[\Re{\{\bar{\chi}(\vec{r}-\vec{r}')/\rho\tilde{\alpha}\}}
\nonumber\\&-&\delta^{(3)}(\vec{r}-\vec{r}')\bar{\mathbb{I}}]\cdot\vec{E}^{\omega}_{D}(\vec{r}')\otimes\vec{E}^{\omega*}_{D}(\vec{r})\Bigr\rangle\Bigr\},\label{E1E2coh}
\end{eqnarray}
where the brackets here denote vacuum expectation values.
The first term on the \emph{r.h.s.} of Eq.(\ref{E1E2coh}) contains the coherent power carried by the field directly radiated by the dipole source into the medium
\cite{LaudonJPB}, $W_{D}^{Coh.}=\frac{\pi\omega}{6\hbar}|\vec{p}_{0}|^{2}
\langle|\vec{E}^{\omega}_{D}(\vec{r})|^{2}\rangle$. Its transverse component is the same expression as that for the coherent intensity carried by a propagating light beam whose source sits out of the medium \cite{Lax}.
In the rest of the \emph{r.h.s.} we can identify part of the field emitted at $\vec{r}'$ by the induced dipoles sitting around the dipole source which propagates towards the source itself located at $\vec{r}$,
\begin{equation}
\vec{E}_{I}^{\omega}(\vec{r})=\int\textrm{d}^{3}r'[\bar{\chi}(\vec{r}-\vec{r}')/\rho\tilde{\alpha}
-\delta^{(3)}(\vec{r}-\vec{r}')\bar{\mathbb{I}}]\cdot\vec{E}^{\omega}_{D}(\vec{r}').
\end{equation}
The subscript $I$ stands both for \emph{induced} and for \emph{indirect}.
Therefore, we can write
\begin{equation}\label{Wcohy}
W^{Coh.}_{\omega}=\frac{\pi\omega}{6\hbar}|\vec{p}_{0}|^{2}
\Bigl[\langle|\vec{E}^{\omega}_{D}(\vec{r})|^{2}\rangle+
\Re{\{\langle\vec{E}^{\omega}_{I}(\vec{r})\cdot\vec{E}^{\omega*}_{D}(\vec{r})\rangle\}}\Bigr]
\end{equation}
and $W^{Coh.}_{\omega}=W_{D}^{Coh.}+W_{I}^{Coh.}$, with
\begin{equation}\label{Wcohyy}
W_{I}^{Coh.}=\frac{\pi\omega}{6\hbar}|\vec{p}_{0}|^{2}
\Re{\{\langle\vec{E}^{\omega}_{I}(\vec{r})\cdot\vec{E}^{\omega*}_{D}(\vec{r})\rangle\}}.
\end{equation}
As expected, Eq.(\ref{Wcohyy}) is the coherent power of that component of the averaged-field induced on the surrounding dipoles which is \emph{in-phase} with Dyson's --see \cite{Remietal} for an analogous computation in a simpler scenario.\\
\indent Alternatively we can write,
\begin{eqnarray}
W_{D}^{Coh.}&=&\frac{-\omega^{3}}{6c^{2}\epsilon_{0}}|\vec{p}_{0}|^{2}\textrm{Tr}\Bigl\{
\Im{\{\bar{G}(\vec{r}',\vec{r})\}}\Bigr\},\label{Wcoher2}\\
W_{I}^{Coh.}&=&\frac{-\omega^{3}}{6c^{2}\epsilon_{0}}|\vec{p}_{0}|^{2}\int \textrm{d}^{3}r'\textrm{Tr}\Bigl\{\Re{}\{\bar{\chi}(\vec{r}-\vec{r}')/\rho\tilde{\alpha}\nonumber\\
&-&\delta^{(3)}(\vec{r}-\vec{r}')\bar{\mathbb{I}}]\}\cdot
\Im{\{\bar{G}(\vec{r}',\vec{r})\}}\Bigr\}.\label{Wincoher2}
\end{eqnarray}
In the diagrammatic representation of $W_{D}^{Coh.}$, $\bar{G}$ is directly attached to the emitter. $\Im{\{G_{\perp}\}}$ yields the transverse normal modes or actual photons which mediate the radiative energy transfer from the emitter into the medium \cite{Andrews}. The diagrammatic representation of $W_{I}^{Coh.}$ can be split in two pieces. The induction piece, $\bar{\chi}(\vec{r}-\vec{r}')/\rho\tilde{\alpha}
-\delta^{(3)}(\vec{r}-\vec{r}')\bar{\mathbb{I}}$, is attached to the emitter. Its non-trivial real part, $\Re{\{\bar{\chi}(\vec{r}-\vec{r}')/\rho\tilde{\alpha}\}}$, does not possess any normal transverse mode. In particular, no actual photon emerges from the emitter, being the induction due to non-radiative energy transfer mediated by virtual photons. The second piece amounts to $\bar{G}$ and is attached to the host dipoles. $\Im{\{G_{\perp}\}}$ yields the transverse normal modes or actual photons emitted by the induced dipoles which are in phase with those of $W_{D}^{Coh.}$.
We conclude that only the spectrum of the power transferred radiatively and directly into the medium is equivalent to the spectrum of external light.\\
\indent In the extinguished emission the factors $\Im{\{\frac{\chi_{\perp,\parallel}(q)}{\rho\tilde{\alpha}}\}}$ of Eqs.(\ref{Gammapincoh},\ref{Gammalincoh}) contain both absorbtion and dispersion, which are in principle indistinguishable observationally. The former is attributed to the intrinsic imaginary part of the bare polarizabilities, $\Im{\{\alpha_{0}\}}$. The latter is captured by the bare transverse modes affected by correlations \cite{BulloughHynne}. That is \emph{incoherent} radiation (in the classical sense) which cannot be written in a form analogous to that of Eq.(\ref{Wcohyy}).\\
\indent Observationally, the power collected in the far field at a distance $r'\gg\xi$ from the emitter --$\xi$ being the typical correlation length-- using an integrating sphere would be the sum of the coherent and the dispersed emissions. On the contrary, the coherent component would be proportional to the integral of the square of the averaged-field measured by an observer on a sphere of radius $r'$. In an effective medium it relates to $W^{Coh.}_{\perp}$ through the Beer-Lambert attenuation factor\footnote{Note that in Eqs.(\ref{Disiclass}-\ref{Wcohyy}) the electric fields are operators acting on the vacuum state. In Eq.(\ref{Wr}) the electric field is an expectation value. The ensemble-average is performed over that expectation value.},
\begin{eqnarray}\label{Wr}
W^{Coh.}(r')&=&W^{Coh.}_{\perp}\exp{\{-2\kappa\omega r'/c\}}\nonumber\\&=&\frac{cn\epsilon_{0}r^{'2}}{2}\int\Bigl|\Bigl\langle\vec{E}(r',\Omega)\Bigr\rangle\Bigr|^{2}\textrm{d}\Omega,
\end{eqnarray}
where $n,\kappa$ are the refractive index and the extinction coefficient of the medium respectively and the big angular brackets denote the ensemble-average.
\section{Relation between light propagation, dipole emission and their vacua}\label{Vacuum}

\subsection{The vacuum of light vs. the vacuum of emission}
\begin{figure}[h]
\includegraphics[height=5.7cm,width=8.9cm,clip]{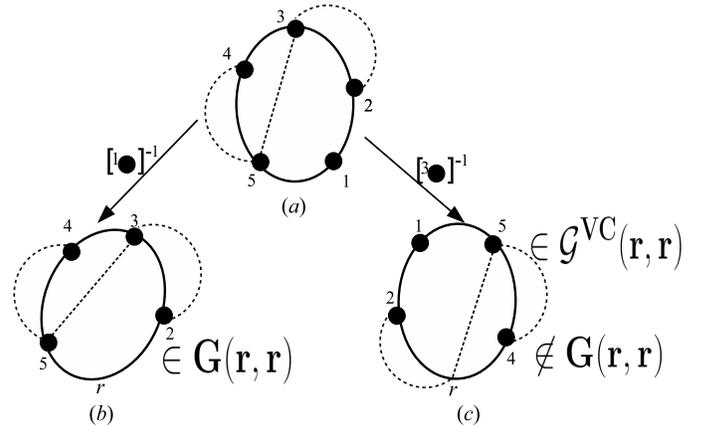}
\caption{($a$) Five-scattering closed-loop with undefined reference frame. ($b$) Fixing the reference frame at the position of the uncorrelated (removed) scatterer labeled as $1$ yields a polariton in $\bar{G}(\vec{r},\vec{r})$. It can be excited either by external light or by dipole emission. ($c$)  Fixing the reference frame at the position of the correlated (removed) scatterer labeled as $3$ yields a polariton in $\bar{\mathcal{G}}^{VC}(\vec{r},\vec{r})$ which is not in $\bar{G}(\vec{r},\vec{r})$. It can be excited by dipole emission only.}\label{fig4}
\end{figure}

It is plain from the above equations that only in free space $\bar{G}(\vec{r},\vec{r})$ and $\bar{\mathcal{G}}(\vec{r},\vec{r})$ are equal to  $\bar{G}^{(0)}(\vec{r},\vec{r})$ and so are
$\mathcal{N}^{light}(\omega)$ and $\mathcal{N}^{emis.}(\omega)$ to $\mathcal{N}^{0}(\omega)=\frac{\omega^{2}}{\pi^{2}c^{3}}$. On the other hand, the fluctuation-dissipation theorem relates the quadratic vacuum fluctuations of the electric field operator at a point $\vec{r}$ with the imaginary part of its propagator from $\vec{r}$ to $\vec{r}$ \cite{Economou,Abriko,Landau},
\begin{equation}\label{Disip}
\langle \Omega|\hat{\vec{E}}_{\omega}(\vec{r})\otimes\hat{\vec{E}}^{\dagger}_{\omega}(\vec{r})|\Omega\rangle=
-\frac{\hbar\omega^{2}}{\epsilon_{0}c^{2}\pi}\Im{\{\bar{\mathbb{G}}(\vec{r},\vec{r};\omega)\}}.
\end{equation}
Therefore, we infer that the photons emitted by an isolated excited dipole propagate in the same vacuum as the radiation which propagates from an external source in absolute absence of dipoles. This is obvious since the source is an isolated dipole itself. However, because in a complex medium
$\bar{G}(\vec{r},\vec{r})\neq\bar{\mathcal{G}}(\vec{r},\vec{r})$, we infer that the EM vacuum in which external light propagates, $|\Omega\rangle^{light}$, is different to that of the photons emitted by the point dipoles, $|\Omega\rangle^{emis.}$,
\begin{eqnarray}
^{light}\langle \Omega|\hat{\vec{E}}_{\omega}(\vec{r})\otimes\hat{\vec{E}}^{\dagger}_{\omega}(\vec{r})|\Omega\rangle^{light}=
\frac{-\hbar\omega^{2}}{\epsilon_{0}c^{2}\pi}\Im{\{\bar{G}_{\perp}(\vec{r},\vec{r};\omega)\}},\label{sl}\\
^{emis.}\langle \Omega|\hat{\vec{E}}_{\omega}(\vec{r})\otimes\hat{\vec{E}}^{\dagger}_{\omega}(\vec{r})|\Omega\rangle^{emis.}=
\frac{-\hbar\omega^{2}}{\epsilon_{0}c^{2}\pi}\Im{\{\bar{\mathcal{G}}(\vec{r},\vec{r};\omega)\}}\label{sp}.
\end{eqnarray}
\indent Let us consider the s.s. virtual cavity formulae of Subsection \ref{2B} and let us represent polaritons and EM fluctuations in general by closed loops of virtual photons created and annihilated at the location of a probe dipole, $\vec{r}$. We can compute the $n$-scattering loops which amount to $\bar{\mathcal{G}}^{VC}(\vec{r},\vec{r})$ out of $(n+1)$-scattering loops with undefined reference frame --see Fig.\ref{fig4}($a$). To do so, $\vec{r}$ must be chosen amongst the $n+1$ position vectors of the scatterers and we must treat the scatterer chosen as a virtual-probe emitter to be removed. If $\vec{r}$ is uncorrelated \emph{w.r.t.} any other point along the original loop, the resultant polariton amounts to $\bar{G}(\vec{r},\vec{r})$ and hence, to $\bar{\mathcal{G}}^{VC}(\vec{r},\vec{r})$ --see Fig.\ref{fig4}($b$). That is the same situation met by the photons coming from an external light source at infinity. On the contrary, if  $\vec{r}$ is correlated  to any of the remaining dipoles in the original loop, the resultant polariton amounts to $\bar{\mathcal{G}}^{VC}(\vec{r},\vec{r})$ but not to $\bar{G}(\vec{r},\vec{r})$ --Fig.\ref{fig4}($c$). Translation invariance holds statistically in the sense that it is the origin of the probe-dipole reference frame, $\vec{r}$, that can be any point in the medium.\\
\indent The finding that the polaritons in $\bar{G}(\vec{r},\vec{r})$ are all included in $\bar{\mathcal{G}}^{VC}(\vec{r},\vec{r})$ was already obtained through the study of coherent emission. Another way to look at this relation is by considering the symmetries of each vacuum. If the origin $\vec{r}$ of one of the $n$-scattering loops of $\bar{\mathcal{G}}^{VC}(\vec{r},\vec{r})$ is translated to any of the remaining $n$ scatterer position vectors of the original $(n+1)$-scattering loop, the resultant loop is also a polariton that amounts to $\bar{\mathcal{G}}^{VC}(\vec{r},\vec{r})$. On the contrary, if the same transformation is carried out on a given $n$-scattering loop of $\bar{G}(\vec{r},\vec{r})$, there is a chance that the resultant polariton amounts to $\bar{\mathcal{G}}^{VC}(\vec{r},\vec{r})$ but not to $\bar{G}(\vec{r},\vec{r})$. Only those transformations which transport  $\vec{r}$ to another uncorrelated scatterer generate a loop in $\bar{G}(\vec{r},\vec{r})$. The group of transformations which leave $\bar{G}(\vec{r},\vec{r})$ invariant is the symmetry group of $|\Omega\rangle^{light}$ and likewise for $\bar{\mathcal{G}}^{VC}(\vec{r},\vec{r})$ and $|\Omega\rangle^{emis.}$. Therefore, we conclude that the group of symmetry of the light vacuum is included in that of the emission vacuum.

\subsection{Local Field Factors as renormalization functions}
The relation of proportionality between Dyson's and the polarization propagator in Fourier space,
\begin{equation}\label{decomp}
\mathcal{G}_{\perp,\parallel}(q;\omega)=\mathcal{L}_{\perp,\parallel}(q;\omega)\:G_{\perp,\parallel}(q;\omega),
\end{equation}
serves us to define $\mathcal{L}^{VC}_{\perp,\parallel}(q;\omega)=\frac{\chi^{\omega}_{\perp,\parallel}(q)}{\rho\tilde{\alpha}}$ as \emph{the local field factors} (LFFs) of the s.s. virtual cavity scenario. A relation of the same kind in the specific scenario of emission from an Onsager cavity drilled in a continuous dielectric was found by Toma\v{s} \cite{Tomas} and Dung \emph{et al.} \cite{Duang06} for long-wavelength modes.
In the present case the proportionality relationship holds for the whole $q$-spectrum and the result applies to a generic random medium.
We now recognize that the induction process which enters $W^{Coh.}_{I}$ in Section \ref{class} is mediated by LFFs.\\
\indent We discuss here in field-theory terms the proportionality relationship between the spectrum of light and that of coherent emission, which derives from the expression $\Im{\{\mathcal{G}^{Coh.}_{\perp,\parallel}(q;\omega)\}}=\Re{\Bigl\{\frac{\chi^{\omega}_{\perp,\parallel}(q)}{\rho\tilde{\alpha}}
\Bigr\}}\Im{\{G_{\perp,\parallel}(q;\omega)\}}$. From our derivation of the \emph{direct} and \emph{indirect} components of the coherent emission we know that $\Re{\Bigl\{\frac{\chi^{\omega}_{\perp,\parallel}(q)}{\rho\tilde{\alpha}}
\Bigr\}}$  take account of the \emph{actual} polarization due to the closest scatterers around the emitter. An analogous phenomenon takes place with the \emph{virtual} polarization of the vacuum in Quantum Electrodynamics (QED) --see eg.\cite{Yildiz} and Ch.7,10 of \cite{Peskin}. Further integration of  $\Im{\{\mathcal{G}^{Coh.}_{\perp,\parallel}(q;\omega)\}}$ in momenta gives the residues $Z_{\perp,\parallel}^{p.\omega}\equiv \Re{\Bigl\{\frac{\chi^{\omega}_{\perp,\parallel}(k^{nor.}_{\perp,\parallel})}{\rho\tilde{\alpha}}
\Bigr\}}$ corresponding to the on-shell modes propagated through the bulk by $G_{\perp,\parallel}(q;\omega)$. Thus, we can write the propagator of coherent modes in a more familiar way, \footnote{The possibility that $\frac{\chi^{\omega}_{\perp,\parallel}}{\rho\tilde{\alpha}}$ present singularities is disregarded here. We do not yet understand their possible physical meaning. It is our guess that it could signal the transition to strong coupling.}
\begin{equation}\label{Coho}
\mathcal{G}^{Coh.}_{\perp,\parallel}(q;\omega)=Z_{\perp,\parallel}^{p.\omega}G_{\perp,\parallel}(q;\omega).
\end{equation}
In field theory terminology, the functions $Z_{\perp,\parallel}^{p.\omega}$ are renormalization functions.
If we stick to the QED interpretation, the electromagnetic field can be renormalized in order to get rid off the $(Z_{\perp,\parallel}^{p.\omega})^{n}$ factors which arise otherwise in the vacuum expectation values of the products of $2n$ EM averaged-field operators. In doing that, the electric field so renormalized is the Dyson field, $\vec{E}_{D}^{\omega}|_{\perp,\parallel}=[Z_{\perp,\parallel}^{p.\omega}]^{-1/2}\vec{E}^{\omega}_{Coh.}|_{\perp,\parallel}$.\\
\indent It is perhaps worth repeating to avoid confusion that the coherent nature of the dipole field here is classical. It is only the formalism borrowed from QED that is quantum.

\subsection{Coherent emission in a Maxwell-Garnett dielectric}\label{FP}
We illustrate the decomposition of the emission spectrum carried out in Section \ref{class} with the computation of the spectrum of coherent emission in a Maxwell-Garnett (MG) dielectric. By definition, an MG dielectric is a random medium made of well-separated dipoles whose spatial distribution is sufficiently well-described by a two-point correlation function of the form \cite{BulloughHynne,Sentenac},
\begin{equation}
h^{MG}(r)=-f(r-\xi)\simeq\left\{
\begin{array}{ll}
 -1 & \textrm{for $r\lesssim \xi$}\\
 0 & \textrm{for $r\gtrsim\xi$}.
\end{array}\right.\nonumber
\end{equation}
$f(r-\xi)$ is this way a spherical-exclusion-volume correlation function of radius $\xi$. Usual forms are those of a Lennard-Jones potential and a hard-sphere potential. For long wavelength modes, $q\xi\ll1$, and low frequencies, $k\xi\ll1$,  the detailed shape of $f(r-\xi)$ is irrelevant and the quasicrystalline approximation becomes exact in the computation of the effective susceptibility \cite{PRLvanTigg}. Thus, it takes the familiar form,
\begin{equation}\label{MGdiel}
\chi_{MG}\equiv\chi^{MG}_{\perp,\parallel}(q\xi=0)=\frac{\rho\tilde{\alpha}}{1-\frac{1}{3}\rho\tilde{\alpha}}.
\end{equation}
We omit for brevity the explicit dependence on $\omega$.
Beside the long wavelength and the low frequency restrictions, the MG formula disregards an inherent self-correlation term in $h(r)$ which gives rise to recurrent scattering. For off-resonant frequencies Eq.(\ref{MGdiel}) is a good approximation up to $\mathcal{O}[(\rho\tilde{\alpha})^{3}]$. Bearing this in mind, we proceed to compute the LDOS of coherent emission, $\mathcal{N}^{Coh.}$, accessible to a point emitter in an MG dielectric. $\mathcal{N}^{Coh.}$ is the LDOS of the observable emission which adjusts to Eq.(\ref{Wr}). The emitter can be either one of the dipole constituents of the medium or an interstitial dipole \cite{PRLdeVries}. Emission must be stimulated with frequencies far from internal resonances in order to avoid recurrent scattering. The calculation is simple for an MG dielectric. Since there are no longitudinal bulk normal modes in the regime of frequencies considered, coherent modes are given by the transverse MG polarization propagator,
\begin{equation}\label{MGtr}
\mathcal{G}^{MG}_{\perp}(q)=\mathcal{L}_{LL}G^{eff}_{\perp}(q),
\end{equation}
In Eq.(\ref{MGtr}), $\mathcal{L}_{LL}=\frac{\epsilon+2}{3}$ is a Lorentz-Lorenz local field factor which derives from Eq.(\ref{MGdiel}) and
\begin{equation}\label{Deff}
G^{eff}_{\perp}(q)=[\epsilon k^{2}-q^{2}]^{-1}
\end{equation}
is the transverse Dyson propagator of an effective medium with dielectric constant $\epsilon$. Straightforward application of Eqs.(\ref{LDOSF},\ref{Coho},\ref{MGtr}) yields,
\begin{equation}\label{MGtr2}
\mathcal{N}^{Coh.}_{MG}(\vec{r})=\frac{\omega^{2}}{\pi^{2}c^{3}}\frac{\Re{\{\epsilon\}}+2}{3}\Re{\{\sqrt{\epsilon}\}}.
\end{equation}
For a non-absorptive medium the index of refraction satisfies $n^{2}=\epsilon$ and thus $\mathcal{N}^{Coh.}_{MG}(\vec{r})\propto(n^{3}+2n)/3$. Note that this dependence on $n$ differs from that of the light spectrum in an effective medium (readily derivable from Eq.(\ref{Deff})), $\mathcal{N}^{light}\sim n$ due to the presence of a local field factor. The precise form of this relation is however model-dependent. As an example, emission from an Onsager-B\"{o}ttcher (OB) cavity would yield $\mathcal{N}^{Coh.}_{OB}(\vec{r})\propto \frac{3n^{3}}{2n^{2}+1}$ instead \cite{Duang06}, where the OB local field factor reads $\mathcal{L}_{OB}=\frac{3\epsilon}{2\epsilon+1}$.\\
\indent From Eqs.(\ref{MGtr}-\ref{MGtr2}) it seems that long-wavelength radiative modes are given only by $\Im{\{G_{\perp}^{eff}(q)\}}$ in the transverse coherent emission. However, $\Im{\{\frac{\chi_{\parallel}(q)}{\rho\tilde{\alpha}}\}}$ contains also transverse parts of the in-free-space propagator \cite{BulloughHynne}.
It can be shown that, for the long-wavelength modes of the effective MG dielectric, the trace of the scattering longitudinal self-polarization propagator can be written as \cite{Mypaperinpreparation},
\begin{eqnarray}\label{isto1}
\varphi^{MG}_{\parallel q\xi\ll1}&=&\int\frac{\textrm{d}^{3}q}{(2\pi)^{3}}
\Bigl[\frac{\chi_{\parallel}(q)/\rho\tilde{\alpha}}{k^{2}[1+\chi_{\parallel}(q)]}-G_{\parallel}^{(0)}\Bigr]^{MG}_{q\xi\ll1}\nonumber\\
&\simeq&2\int\frac{\textrm{d}^{3}q}{(2\pi)^{3}}\mathcal{L}_{LL}G^{eff}_{\perp}(q)[\mathcal{L}_{LL}-1].
\end{eqnarray}
As a result, the total LDOS for far field emission contains a dispersive component proportional to $\mathcal{L}_{LL,OB}(\mathcal{L}_{LL,OB}-1)n$ in the MG and the OB models respectively. We can conclude that generically and for the far field emission in an effective medium,  one local field factor enters the coherent spectrum while two factors enter the total emission spectrum. Only the latter is well-known \cite{Juzeliunas}.

\section{The Vacuum Energy}\label{Casimir}
It has been recognized for a long time that the Casimir effects, the van der Waals (vdW) forces and the Lamb-shift share a common origin. It is customary to ascribe the vdW and Lamb energies to short-range interactions while the Casimir effects are thought of as long-range interactions.
According to this qualitative distinction the vdW and Lamb energies have a microscopical origin. It roots on the EM interactions between the dipole constituents of the dielectric. For instance, the sum of many-body interactions yields a sort of virial expansion for the vdW energy \cite{FelderhofvdW}. On the contrary, the Casimir energy would have a macroscopical origin. It roots on the long-range (retarded) EM interactions between macroscopic dielectrics. Typically only the long-wavelengths of the EM field
are relevant for the Casimir energy and their spectrum gets discrete as the field obeys macroscopic boundary conditions. Some authors refer to the energy of discrete modes as Casimir  energy while they term bulk energy that of the continuous spectrum \cite{Milonnietal,Molina,Milton}.\\
\indent Rather than making the distinction on the basis of the interaction range, an alternative terminology consists of referring to the ground state EM energy of the system  when stored in the material degrees of freedom (d.o.f.) of the dielectric as binding energy \cite{Abriko,Obada,FelderhofvdW}. On the contrary, when the ground state EM energy is computed out of the EM vacuum fluctuations it is termed vacuum energy \cite{Boyer}. Evidently, this distinction is just conceptual and the total binding energy must equal the total vacuum energy. In either case, the EM energy manifests in the shifts of bound state levels or in the shifts of the resonant frequencies of optical normal modes.\\
\indent Since we deal with an infinite volume complex medium, surface terms and discrete modes are absent and there is no need for using the interaction-range-based classification. The computation of $\mathcal{G}^{VC}$ and the spectrum of vacuum fluctuations allows us to compute the total vacuum energy density, $\mathcal{F}^{V}$. On the other hand, the renormalization of the single particle polarizabilities allows us to identify the shifts in the binding energy of atomic states. Without ambiguity we can term the energy shift of atomic levels Lamb-shift \cite{Wylie,WelschI}. Correspondingly, the energy density stored in the internal d.o.f. of every atom is the Lamb energy density, $\mathcal{F}^{L}$. It follows from the above reasoning that   $\mathcal{F}^{L}$ must be part of $\mathcal{F}^{V}$. Nonetheless, because other than internal d.o.f. are present in a molecular dielectric, it is expected that the binding energies of collective (i.e. cluster) d.o.f contribute also to $\mathcal{F}^{V}$. In references \cite{Schaden,Milonnietal} the authors have found that, al leading order in $\rho\alpha$, the Lamb-shift can be derived from the total variation of the vacuum energy. In turn this would imply that, at leading order, all the vacuum energy gets stored in the internal d.o.f. of the dipole constituents. However, that result is not totally conclusive as near field interactions and local field factors are neglected there.\\
\indent The aims of this Section are, in the first place to obtain an analytical formula for $\mathcal{F}^{V}$ which facilitates future investigations about its physical content. Second, to compute the vacuum energy of an effective medium to estimate in first approximation the contribution due to local field factors.

\subsection{The vacuum energy of a homogeneous and isotropic molecular dielectric}
We apply the variational method of Schwinger to the computation of the total vacuum energy density of a generic complex medium made of point dipoles. According to that approach the vacuum energy emerges as an effective potential of the EM field. This effective potential is induced as the dipole constituents are brought adiabatically from infinity and are assembled until completing the final dielectric configuration. To an infinitesimal variation on the dielectric properties of the medium corresponds a variation on the spectrum of vacuum fluctuations, $\mathcal{N}^{emis.}$, and hence a variation on the effective potential. The integration of the latter variations is $\mathcal{F}^{V}$. Following \cite{SchwingerMath,Schwinger4095,Schwinger2105},
the Schwinger formula for the effective action reads,
\begin{equation}
\mathcal{S}[\bar{\mathcal{H}}]=-i\frac{\hbar}{2}\textrm{Tr}\Bigl\{\int_{0}^{\infty}\frac{\textrm{d}u}{u}e^{iu\bar{\mathcal{H}}+i\eta}\Bigr\},\quad\eta\rightarrow0^{+}
\end{equation}
where the lower limit of integration can be set to zero after regularization of divergent terms and $\bar{\mathcal{H}}$ is the proper-time Hamiltonian operator. In our case $\bar{\mathcal{H}}$  is the inverse of the self-polarization propagator, $\bar{\mathcal{H}}=\bar{\mathcal{G}}^{-1}$, in accordance to the fluctuation-dissipation theorem \cite{MilonniPRA}. The Casimir energy relates to $\mathcal{S}$  by $\mathcal{F}^{V}=-\Re{\{\mathcal{S}\}}/(\mathcal{V}\mathcal{T})$, $\mathcal{T}$ being an asymptotic time of observation and $\mathcal{V}$ being the (infinite) volume occupied by the dielectric material. This way, variations on $\mathcal{F}^{V}$ are due to variations of the self-polarization propagator.  Replacing $\bar{\mathcal{H}}$ with $\bar{\mathcal{G}}^{-1}$, performing the variational derivative, integrating in $u$ and taking the trace, one obtains,
\begin{eqnarray}\label{SchwingerGR}
\delta\mathcal{F}^{V}&=&\frac{\hbar}{2}\Im{}\Bigl\{\int_{-\infty}^{\infty}\frac{\textrm{d}\omega}{(2\pi)} \int\frac{\textrm{d}^{3}q}{(2\pi)^{3}}\nonumber\\&\times&
[2\mathcal{G}_{\perp}(q)\delta\mathcal{G}^{-1}_{\perp}(q)+\mathcal{G}_{\parallel}(q)\delta\mathcal{G}^{-1}_{\parallel}(q)]\Bigr\}.
\end{eqnarray}
The variational derivatives can be performed \emph{w.r.t.} any functional that characterizes the initial and final vacuum state. In our case, the final state is determined by $\bar{\mathcal{G}}^{VC}$ which is function of $\bar{G}$, $\bar{\chi}$ and $\tilde{\alpha}$. The initial state is that in which the EM field is free and the dipoles do not interact, being decoupled from the radiation field. For the two-level atom model of Appendix \ref{appendy} their 'initial' polarizability is $\alpha'=\frac{\alpha_{0}}{3}/(1-\omega^{2}/\omega_{0}^{2})$. Functional integration of Eq.(\ref{SchwingerGR}) yields,
\begin{eqnarray}
\mathcal{F}^{V}&=&\frac{-\hbar}{2}\Im{}\Bigl\{\int_{-\infty}^{\infty}\frac{\textrm{d}\omega}{2\pi} \int\frac{\textrm{d}^{3}q}{(2\pi)^{3}}\ln{[(\alpha'/\tilde{\alpha})^{3}]}\Bigr\}\label{Flamb}\\
&+&\frac{-\hbar}{2}\Im{}\Bigl\{\int_{-\infty}^{\infty}\frac{\textrm{d}\omega}{2\pi} \int\frac{\textrm{d}^{3}q}{(2\pi)^{3}}\ln{}\Bigl[\chi^{2}_{\perp}G^{2}_{\perp}\chi_{\parallel}G_{\parallel}\Bigr]\Bigr\}\label{bulky}\\
&-&\frac{-\hbar}{2}\Im{}\Bigl\{\int_{-\infty}^{\infty}\frac{\textrm{d}\omega}{2\pi} \int\frac{\textrm{d}^{3}q}{(2\pi)^{3}}\ln{}\Bigl[[G^{(0)}_{\perp}]^{2}\Bigr]\Bigr\},\label{lacagada}
\end{eqnarray}
where the $\omega,q$-dependence of the integrands has been omitted.
An expression like this was firstly obtained by Bullough and Obada for a cubic molecular crystal \cite{Obada}. The authors however failed in not considering the renormalization of the single-particle polarizability. This is crucial in order to identify the Lamb energy. For future purposes we will refer to the terms which involve the bare and Dyson's transverse propagators in Eqs.(\ref{bulky},\ref{lacagada}) as Schwinger-bulk energy,
\begin{equation}\label{schw}
\mathcal{F}^{Sch.}_{bulk}=\frac{-\hbar}{2}\Im{}\Bigl\{\int_{-\infty}^{\infty}\frac{\textrm{d}\omega}{2\pi} \int\frac{\textrm{d}^{3}q}{(2\pi)^{3}}\ln{}\Bigl[G^{2}_{\perp}/[G^{(0)}_{\perp}]^{2}\Bigr]\Bigr\}.
\end{equation}
\indent Beside the in-free-space Lamb-shift Buhmann \emph{et al.}  have computed the additional shift due to the interaction of every dipole with the rest of the dielectric, $\mathcal{E}^{Lsh}_{sc}$ \cite{WelschI,WelschII}. In other words, $\mathcal{E}^{Lsh}_{sc}$ amounts to the energy stored in the atomic d.o.f. that it would cost to take out of the medium one of the dipoles -- note that other possible energy costs in the removal process are not accounted for in $\mathcal{E}^{Lsh}_{sc}$. $\mathcal{E}^{Lsh}_{sc}$ is made of a resonant term which only contributes if the dipole is in an excited state, $\mathcal{E}^{Lsh}_{res.}$. And an off-resonant term which depends on the polarizability of each atomic state, $\mathcal{E}^{Lsh}_{off}$. We just adapt the computations in \cite{WelschI,WelschII} to our results in Sections \ref{2B} and \ref{polarizability} for two-level atoms in the VC scenario and obtain,
\begin{eqnarray}
\mathcal{E}^{Lsh}_{res.}&=&\frac{|\mu|^{2}}{3\epsilon_{0}}
\Re{\{k^{2}[2\varphi^{VC}_{\perp}+\varphi^{VC}_{\parallel}]|_{k=k_{res}+ic\Gamma}\}},\label{ee3}\\
\mathcal{E}^{Lsh}_{off}&=&-\frac{\hbar c}{4\pi}\int_{0}^{\infty}\textrm{d}u\:u^{2}[2\varphi^{VC}_{\perp}+\varphi^{VC}_{\parallel}]|_{k=iu}\nonumber\\
&\times&[\tilde{\alpha}|_{k=iu}+\tilde{\alpha}|_{k=-iu}],\label{eer3}
\end{eqnarray}
where the $\varphi$-factors are implicit functions of $k$. Eq.(\ref{ee3})  equals $c\hbar(k_{res}-k_{0})$ in Eq.(\ref{kres}) except for the fact that it is not evaluated at $k=k_{res}$ but at $k=k_{res}+ic\Gamma$ instead. For the same polarizability model,  $\tilde{\alpha}|_{k=\pm iu}$ in Eq.(\ref{eer3}) is given by Eq.(\ref{Lorentzian}) evaluated at the imaginary frequency $ck=\pm icu$. The expansion of the term in Eq.(\ref{Flamb}) allows to verify that both the in-free-space Lamb-shift and $\rho\mathcal{E}^{Lsh}_{sc}$ enter that equation but with opposite sign\footnote{The proof of the equality $Eq.(\ref{Flamb})=-3\mathcal{F}^{L}$ will be given somewhere else.}. This leads to an interpretation of  Eqs.(\ref{Flamb},\ref{bulky},\ref{lacagada}) slightly different to that in \cite{Obada}. That is, $\mathcal{F}^{V}$ takes account of the in-free-space EM modes of Eq.(\ref{lacagada}) and the Lamb energy of Eq.(\ref{Flamb}) and substitutes them with the optical modes and/or binding energies of the coupled system of Eq.(\ref{bulky}).\\
\indent The decomposition of $\mathcal{F}^{V}$ suggests two questions. Firstly, in view of the above preliminary results, it seems likely that $\mathcal{F}^{L}$ is different from $\mathcal{F}^{V}$. Consequently, the total energy cost in the removal of one of the dipoles from the dielectric would be different to  $\mathcal{E}^{Lsh}_{sc}$. This must be verified. Second, provided that $\mathcal{F}^{L}\neq\mathcal{F}^{V}$, it should be investigated whether or not there exists any observational signature of such a distinction. We leave these issues for future work.

\subsection{The vacuum energy of an infinite-volume effective medium}
In this section we abandon the microscopical description of the dielectric and concentrate on the long wave length modes of the vacuum energy density in an infinite-volume effective dielectric. The original computation carried out by Schwinger \cite{Schwinger2105} was motivated by the problem of the sonoluminescence phenomenon on big bubbles. In here we will not try to address in detail this problem since it involves secondary issues like those of surface corrections \cite{Molina}, etc. Our purpose is to investigate how the presence of local field factors, ignored in previous approaches, modifies the usual Schwinger-bulk energy.\\
\indent The Schwinger bulk scenario restricts to low frequencies and long-wavelength modes in comparison to the typical length scale of the dielectric microstructure. Thus, without much loss of generality we adopt the Maxwell-Garnett model of Subsection \ref{FP}. This way, the MG relation between $\epsilon$ and $\tilde{\alpha}$ in Eq.(\ref{MGdiel}) can be used and the Lamb-shift can be considered integrated already in $\tilde{\alpha}$. Following \cite{Schwinger2105} we take a large dielectric body and ignore the surface terms. Thus, we will refer to the resultant energy as effective-volume energy to distinguish it from the Schwinger-bulk energy and from the total vacuum energy which would include short wavelengths.\\
\indent Instead of using the exact formulae of Eqs.(\ref{Flamb},\ref{bulky},\ref{lacagada}) it is easier here to start with the functional derivatives of  Eq.(\ref{SchwingerGR}). For an effective medium of uniform permittivity $\epsilon'(\omega)=1+\chi^{'}_{MG}(\omega)$ the variations on the \emph{r.h.s.} of Eq.(\ref{SchwingerGR}) stand for
\begin{equation}
\delta\mathcal{G}^{-1}_{\perp,\parallel}(q)=\delta\epsilon'\frac{\delta}{\delta\epsilon'}\{\mathcal{G}^{-1}_{\perp,\parallel}(q)\},
\end{equation}
where the explicit dependence on $\omega$ has been omitted. Using the formulae of Subsection \ref{FP}, we are left with
\begin{eqnarray}
\mathcal{F}^{V}_{MG}&=&\frac{\hbar}{2}\Im{}\Bigl\{\int_{-\infty}^{\infty}\frac{\textrm{d}\omega}{(2\pi)}\int \frac{\textrm{d}^{3}q}{(2\pi)^{3}}\int_{1}^{\epsilon}\delta\epsilon'\nonumber\\&\times&
\Bigl(2\mathcal{L}_{LL}G^{eff}_{\perp}(q)\frac{\delta}{\delta\epsilon'}\{\mathcal{L}^{-1}_{LL}[G^{eff}_{\perp}(q)]^{-1}\}\nonumber\\
&+&\mathcal{L}_{LL}G^{eff}_{\parallel}\frac{\delta}{\delta\epsilon'}\{\mathcal{L}^{-1}_{LL}[G^{eff}_{\parallel}]^{-1}\}\Bigr)\Bigr\},\nonumber
\end{eqnarray}
where $G^{eff}_{\parallel}=\frac{1}{\epsilon' k^{2}}$ and we have omitted the constant term $-3\hbar\Im{}\Bigl\{\int_{0}^{\infty}\frac{\textrm{d}\omega}{2\pi} \ln{[\alpha']}\Bigr\}$ which is a reminder of the internal atomic binding energy of the isolated dipoles. The above formula presents two separable contributions,
\begin{eqnarray}
\mathcal{F}^{V}_{MG}&=&2\hbar\int_{0}^{\infty}\frac{\textrm{d}\omega}{(2\pi)}\frac{\omega^{2}}{c^{2}}\Im{\Bigl\{\int_{1}^{\epsilon}\delta\epsilon'\int\frac{\textrm{d}^{3}q}{(2\pi)^{3}}
G_{\perp}^{eff}\Bigr\}}\label{Sch}\\
&-&\hbar\int_{0}^{\infty}\frac{\textrm{d}\omega}{(2\pi)}\int\frac{\textrm{d}^{3}q}{(2\pi)^{3}}\Im{\{\ln{[\mathcal{L}_{LL}^{3}/\epsilon]}\}}.\label{LFFterm}
\end{eqnarray}
The term in Eq.(\ref{Sch}) is the usual Schwinger-bulk term equivalent to that in Eq.(\ref{schw}) for an effective medium,
\begin{eqnarray}
\mathcal{F}^{Sch.}_{bulk}&=&\frac{\hbar}{6\pi^{2}c^{3}}\Re{\Bigl\{\int_{0}^{\infty}\textrm{d}\omega\:\omega^{3}[1-\epsilon^{3/2}(\omega)]\Bigr\}}\nonumber\\
&\simeq&\frac{\hbar}{6\pi^{2}c^{3}}\int_{0}^{\infty}\textrm{d}\omega\:\omega^{3}[1-n^{3}(\omega)],\label{Sch1}
\end{eqnarray}
with $n(\omega)=\Re{\{\sqrt{\epsilon(\omega)}\}}$.
Since $\mathcal{F}^{Sch.}_{bulk}$ has been profusely studied in the literature we will not attempt to evaluate it further --see eg. \cite{Milton}. Nonetheless, for future discussions let us write $\mathcal{F}^{Sch.}_{bulk}$ as
\begin{equation}\label{Sch2}
\mathcal{F}^{Sch.}_{bulk}=\int_{0}^{\infty}\textrm{d}\omega\frac{-1}{2}\hbar\omega[\mathcal{N}^{Sch.}_{bulk}(\omega)-\frac{1}{3}\mathcal{N}^{0}(\omega)].
\end{equation}
By equating Eq.(\ref{Sch1}) with Eq.(\ref{Sch2}) we define $\mathcal{N}^{Sch.}_{bulk}(\omega)=\frac{\omega^{2}}{3\pi^{2}c^{3}}n^{3}(\omega)$.\\
\indent The term in Eq.(\ref{LFFterm}) comes from the combination of the local field factors and the effective longitudinal propagator. We will denote it by $\Delta\mathcal{F}_{MG}$. It is not associated to normal modes. As a matter of fact, only longitudinal photons contribute to the LFFs in the MG model. Therefore, $\Delta\mathcal{F}_{MG}$ must contain the contribution of internal resonances. We can estimate its value using a Lorentzian dielectric constant of band-width $\Gamma$ and resonant frequency $\omega_{res}$,
\begin{equation}
\epsilon=1+\frac{f\omega_{res}^{2}}{\omega_{res}^{2}-\omega^{2}-i\omega\Gamma}.\label{lae}
\end{equation}
With the above formula, Eq.(\ref{LFFterm}) reads approximately,
\begin{equation}\label{LFFG}
\Delta\mathcal{F}_{MG}\approx\frac{\rho f^{2}}{12}\Bigl[
\frac{\hbar\omega_{res}}{2}+\frac{\hbar\Gamma}{2\pi}\Bigr],
\end{equation}
where $f$ is an effective oscillator strength with $f\ll1$ and $\rho^{-1}$ is the volume-per-dipole such that $\int\frac{\textrm{d}^{3}q}{(2\pi)^{3}}=\rho$.
 Note that the isolated contribution of LFFs would be of order $f$ instead,
\begin{equation}\label{LFFG2}
\mathcal{F}^{LFF}_{MG}=-\rho\hbar\int_{0}^{\infty}\frac{\textrm{d}\omega}{(2\pi)}\Im{\{\ln{[\mathcal{L}_{LL}^{3}]}\}}\approx\frac{\rho f}{2}\Bigl[
\frac{\hbar\omega_{res}}{2}+\frac{\hbar\Gamma}{2\pi}\Bigr].\nonumber
\end{equation}
\indent Except for numerical prefactors, the expression for $\Delta\mathcal{F}_{MG}$ appears generic for the effective-volume energy density of long-wavelength modes and regardless
of the particular microscopical dielectric model. Nonetheless, some comments are in order regarding the particularities of the MG model.
It was mentioned in Subsection \ref{FP} that the MG model neglects recurrent scattering. However, at resonance scatterers overlap
optically and recurrent scattering becomes dominant. Therefore, deviations from the MG results are expected \cite{Felderhof3}.
Nevertheless, although the precise form of the LFFs changes at resonance, the integration over $\omega$ of Eq.(\ref{LFFterm})
is not expected to vary \emph{w.r.t.} the result of Eq.(\ref{LFFG}) but for the $f$-dependent prefactors.

\section{Discussion on LDOS$_{\textrm{es}}$}\label{disc}

Firstly, let us comment on the distinction between $\mathcal{N}^{Sch.}_{bulk}(\omega)=\frac{\omega^{2}}{3\pi^{2}c^{3}}n^{3}(\omega)$ and $\mathcal{N}^{light}(\omega)=\frac{\omega^{2}}{\pi^{2}c^{3}}n(\omega)$ for an effective medium. Although both spectra are computed out of $G^{eff}_{\perp}$ and thus ascribed to the light vacuum, their respective dependence on the refractive index and their physical meaning are quite different. We read from Eq.(\ref{Sch}) that $\mathcal{N}^{Sch.}_{bulk}\propto n^{3}$ originates from the integration of the variations of $[G_{\perp}^{eff}]^{-1}$, $\delta [G^{eff}_{\perp}(q)]_{\epsilon'}^{-1}=\frac{\omega^{2}}{c^{2}}\delta\epsilon'$, weighted by the light LDOS of each intermediate dielectric configuration of permittivity $\epsilon'$,  $\mathcal{N}^{light}_{\epsilon'}\sim\Im{\{G_{\perp}^{eff}(\vec{r},\vec{r})\}_{\epsilon'}}$. Because the Schwinger-bulk energy can be written as an integral in momenta by identifying $\omega=q\frac{c}{n}$, $\mathcal{F}^{Sch.}_{bulk}=\int\frac{\textrm{d}^{3}q}{(2\pi)^{3}}\hbar cq(1/n-1)$ --eg. \cite{Molina}, the $n^{3}$ dependence of $\mathcal{N}^{Sch.}_{bulk}(\omega)$ can be fairly attributed to its $c^{-3}$ dependence \cite{Schaden}. On the contrary,  $\mathcal{N}^{light}\propto n$ is just the light LDOS of the final configuration of permittivity $\epsilon$ and is simply proportional to $\Im{\{G_{\perp}^{eff}(\vec{r},\vec{r})\}_{\epsilon}}$. Alternatively, the relation $\mathcal{N}^{light}=3n^{-2}\mathcal{N}^{Sch.}_{bulk}(\omega)$ can be obtained if one starts from Fermi's Golden rule and restrict the quadratic vacuum fluctuations to those of the Dyson field. The macroscopic quantization of each field operator yields a factor $n^{-1}$ while the remainder is the sum over momenta proportional to $n^{3}$ \cite{Snoeks}. Note however that we have defined originally the LDOSes through relations of the form of Eq.(\ref{LDOSF}). That is, by definition an LDOS is proportional to the imaginary part of some physically meaningful Green's function. It is in this sense that $\mathcal{N}^{Sch.}_{bulk}$ derives (i.e., is not defined) from the original $\mathcal{N}^{light}$ and likewise for the LDOS of $\mathcal{F}^{V}$ which derives from $\mathcal{N}^{emis.}$.\\
\indent Second, let us emphasize that $\mathcal{N}^{light}$ must not be identified with the total radiative LDOS. As mentioned in Section \ref{polarizability}, both the stimulated and the spontaneous emission of a point dipole are proportional to the quadratic fluctuations of the self-polarization field in $|\Omega\rangle^{emis.}$ \cite{MilonniAmJ,LaudonJPB}. Thus, we can write,
\begin{equation}
\Gamma\propto\: ^{emis.}\langle \Omega|\hat{\vec{E}}^{\omega_{res}}(\vec{r})\cdot\hat{\vec{E}}^{\omega_{res}^{\dagger}}(\vec{r})|\Omega\rangle^{emis.}.
\end{equation}
The above expression together with Eqs.(\ref{sl},\ref{sp}) and Eqs.(\ref{LDOSIper},\ref{LDOSIparal}) show clearly that it is not $\mathcal{N}^{light}$\footnote{$\mathcal{N}^{light}$ is denoted by $N_{rad}$ in \cite{RMPdeVries}. Also in this context, radiative emission is synonym of far field emission.} \emph{the relevant part of the LDOS entering Femi's Golden rule in the descriptions of the rate of spontaneous radiative decay} as erroneously claimed in \cite{RMPdeVries}. This has been appreciated by a number of authors --see eg. \cite{LaudonJPB,PRLdeVries,Tomas,Duang06,Juzeliunas} who have attributed the discrepancy to the presence of LFFs which relate the microscopical field to the macroscopical-Maxwell field. We have proved microscopically that this is indeed the case, but rather than relating microscopical and macroscopical fields, LFFs in Fermi's Golden rule relate the self-polarization propagator to the Dyson-bulk propagator, they both being computed at the same microscopical level. LFFs emerge in Fermi's Golden rule as a result of taking proper account of correlations. We found in Section \ref{class} that the correct statement is that it is only the power transferred radiatively and directly from the emitter into the medium that has the spectrum of external light.
There are other contributions to the total power which, although not transferred directly from the emitter into the medium, are radiative anyways. In particular, the \emph{indirect} coherent emission is of course radiative, although that emission is carried out by induced dipoles and not by the emitter itself. In the far field, at a distance much greater than the typical correlation length between the emitter and the surrounding dipoles, it would not be possible to distinguish between \emph{direct} and \emph{indirect} coherent radiation.\\
\indent The restrictive identification in \cite{RMPdeVries} of the radiative LDOS with $\mathcal{N}^{light}$ seems to have led to confusion on the authors of \cite{RemiMole} who are interested in the relation between the refractive index and the radiative decay rate of an emitter embedded in a random medium. They argue that, since in the long-wavelength limit of the effective medium theory, $q\xi\rightarrow0$, the dielectric constant becomes a local and isotropic function, correlation effects can be disregarded and the effective transverse propagator of Eq.(\ref{Deff}) determines the radiative LDOS which enters Fermi's Golden rule. On the contrary, because under no physically acceptable condition $\mathcal{L}|_{q\xi\rightarrow0}=1$ holds --eg. $\mathcal{L}_{LL}$, that argument is generically erroneous.\\
\indent A similar reasoning applies to \cite{Tip} where the authors study the radiative decay of a dipole emitter embedded in a piecewise continuous dielectric. They take as LDOS of reference that of a homogeneous (i.e. one-piece) medium, $\sim n$. However that is not the appropriate LDOS for emission but for light propagation. The reason being that it is not possible to embed an emitter in a continuous medium and take a continuous limit such that the emitter forms part of the medium and the effects of LFFs disappear -- eg. in the limit $Rq\rightarrow0$ of an Onsager cavity of radius $R$, $\mathcal{L}_{Rq\rightarrow0}=\mathcal{L}_{OB}\neq1$ \cite{Tomas,Duang06}. Otherwise, unphysical divergences appear for the case that the permittivity be a complex number. In conclusion, while $\mathcal{N}^{light}$ is well defined in a strictly homogeneous medium, $\mathcal{N}^{emis.}$ is not. The computation of $\mathcal{G}_{\perp}$ in \cite{Tip} for an heterogeneous medium is correct provided that the dielectric function takes proper account of the emitter embedding. However, in such a case there are no divergences in $\Im{\{\mathcal{G}_{\parallel}\}}$ and there is no need to restrict the radiative LDOS to $\Im{\{\mathcal{G}_{\perp}\}}$ as claimed by the authors. In particular, the statement that $\Im{\{\textrm{Tr}\{\bar{\mathcal{G}}\}\}}=2\Im{\{\mathcal{G}_{\perp}\}}$ in a medium of real dielectric constant is in general incorrect. If properly computed, $2\Im{\{\mathcal{G}_{\perp}\}}$ in an effective medium yields only coherent modes.\\
\indent Note however that if one were interested only in detecting the band gaps of radiative emission in an effective medium, the detection of the gaps of $\mathcal{N}^{light}$, which correspond to bands of total reflectivity, might suffice. The reason is that, for an effective medium like that of an MG dielectric, the total radiative spectrum is given by $\mathcal{L}^{2}_{q\xi=0}(\omega)\mathcal{N}^{light}(\omega)\propto\mathcal{L}^{2}_{q\xi=0}(\omega)n(\omega)$ \cite{PRLdeVries,Duang06,Juzeliunas,Mypaperinpreparation}. Therefore, unless the LFFs diverge at those frequencies at which the index of refraction vanishes, the gaps on $\mathcal{N}^{light}(\omega)$ coincide with those of the spectrum of total radiation. It is to this respect that the work of \cite{RemiMole} is still useful. In physical terms, this implies that there does not exist coupling of either the emitter itself or the surrounding induced dipoles to propagating modes. This however does not preclude the non-radiative energy transfer between the emitter and the induced dipoles.\\
\indent Finally, a comment is in order on the application of Eqs.(\ref{LDOSIper}-\ref{Xi}) to the spontaneous decay rate of a dipole in a statistically homogeneous complex medium.
When the emission is stimulated, under stationary conditions the rate at which the dipole absorbs the energy supplied by the external source is (ideally) balanced by the rate at which the dipole radiates that energy. Therefore, its polarizability can be considered at any time equivalent to that of the rest of dipoles in the medium and Eqs.(\ref{LDOSIper}-\ref{Xi})
are strictly applicable so that $\mathcal{N}^{emis.},W_{\omega}\propto\textrm{Tr}\Bigl\{\Im{\{\bar{\mathcal{G}}^{VC}(\vec{r},\vec{r};\omega)\}}\Bigr\}$.
In contrast, when the dipole decays spontaneously from an excited state, its polarizability is in general different to that of the rest of surrounding
dipoles. This brakes manifestly the statistical translation-invariance of the medium at the emitter location since it behaves as an impurity. As a consequence,  Eqs.(\ref{LDOSIper}-\ref{Xi})
may be applicable only in an approximate manner. This is at the root of the distinction between the virtual cavity and the real cavity scenarios \cite{PRLdeVries,Topsying}.

\section{The propagator of the self-polarization field in numerical simulations}\label{T}
The computation of $\tilde{\alpha}$ and the measurement of all the components of $W_{\omega}$ can be simulated numerically using the so-called Coupled-Dipole-Method (CDM) \cite{Draine}. In that method, the transference matrix $\bar{\mathrm{t}}$ of Subsection \ref{light} is computed for each configuration of point dipoles before performing the average over the configuration ensemble. Let us take one of those configurations composed by $N+1$ host scatterers. We decide to stimulate the dipole at $\vec{R}_{0}$ with an incident (inc) monochromatic field of frequency $\omega=kc$, $\vec{E}_{0}$. That is, $\vec{E}_{inc}(\vec{r})=\vec{E}_{0}$ if $\vec{r}=\vec{R}_{0}$ and $\vec{E}_{inc}(\vec{r})=\vec{0}$ otherwise. Generally the dipoles are not all equivalent and have different renormalized polarizabilities $\tilde{\alpha}^{i}$, $i=0,..,N$. Therefore, the dipole moment of the emitter reads in terms of the incident field,
\begin{equation}\label{pemuto}
\vec{p}(\vec{R}_{0})=\epsilon_{0}\tilde{\alpha}^{0}\vec{E}_{0}.
\end{equation}
On the other hand, the total field at the emitter location is the sum of the incident field plus the self-polarization field which the dipole moment $\vec{p}(\vec{R}_{0})$ itself creates at $\vec{R}_{0}$. That is,
\begin{eqnarray}\label{Eemit}
\vec{E}(\vec{R}_{0})&=&\vec{E}_{0}-\frac{k^{2}}{\epsilon_{0}}\bar{\mathfrak{g}}(\vec{R}_{0},\vec{R}_{0})\cdot\vec{p}(\vec{R}_{0})
\nonumber\\&=&\vec{E}_{0}-k^{2}\tilde{\alpha}^{0}\bar{\mathfrak{g}}(\vec{R}_{0},\vec{R}_{0})\cdot\vec{E}_{0},
\end{eqnarray}
where $\bar{\mathfrak{g}}(\vec{R}_{0},\vec{R}_{0})$ is the self-polarization propagator of the $0^{th}$ dipole for the scatterer configuration considered. The $\omega$-dependence is omitted hereafter. Our interest is in the second term on the \emph{r.h.s.} of Eq.(\ref{Eemit}). It can be written also as the sum of all the fields which propagate from each of the induced dipoles to $\vec{R}_{0}$,
\begin{equation}\label{pemit}
-k^{2}\tilde{\alpha}^{0}\bar{\mathfrak{g}}(\vec{R}_{0},\vec{R}_{0})\cdot\vec{E}_{0}=\frac{-k^{2}}{\epsilon_{0}}
\sum_{i=0}^{N}\bar{G}^{(0)}(\vec{R}_{0}-\vec{R}_{i})\cdot\vec{p}(\vec{R}_{i}).
\end{equation}
Except for the emitter, the rest of dipoles are only induced by their mutual interactions. The dipole moments induced can be computed formally using the $\bar{\mathrm{t}}$-matrix of the system. $\bar{\mathrm{t}}(\vec{R}_{i},\vec{R}_{j})$ yields the dipole induced at some point $\vec{R}_{i}$ as a result of its interaction with the dipole excited by the incident field at some point $\vec{R}_{j}$. The dipole moment of a dipole located at $\vec{R}_{i}$ reads,
\begin{equation}
\vec{p}(\vec{R}_{i})=-\Bigl(\frac{k^{2}}{\epsilon_{0}}\Bigr)^{-1}\sum_{j=0}^{N}\bar{\mathrm{t}}(\vec{R}_{i},\vec{R}_{j})\cdot\vec{E}_{inc}(\vec{R}_{j}).
\end{equation}
Because in our case the only dipole excited externally is that at $\vec{R}_{0}$, we get
\begin{equation}\label{lapi}
\vec{p}(\vec{R}_{i})=-\Bigl(\frac{k^{2}}{\epsilon_{0}}\Bigr)^{-1}\bar{\mathrm{t}}(\vec{R}_{i},\vec{R}_{0})\cdot\vec{E}_{0}.
\end{equation}
Inserting Eq.(\ref{lapi}) into Eq.(\ref{pemit}) we end up with,
\begin{equation}\label{Gdiscr}
\bar{\mathfrak{g}}(\vec{R}_{0},\vec{R}_{0})=\frac{1}{-k^{2}\tilde{\alpha}^{0}}\sum_{i=0}^{N}
\bar{G}^{(0)}(\vec{R}_{0}-\vec{R}_{i})\cdot\bar{\mathrm{t}}(\vec{R}_{i},\vec{R}_{0}).
\end{equation}
Should all the dipoles be equivalent with $\tilde{\alpha}^{i}=\tilde{\alpha}$ $\forall i$ and density $\rho$, it is immediate to obtain
 $\Bigl<\bar{\mathfrak{g}}(\vec{R}_{0},\vec{R}_{0})\Bigr>=\bar{\mathcal{G}}^{VC}(\vec{R}_{0},\vec{R}_{0})$ as given in Eq.(\ref{infunctionofT}).\\

\indent A comment is in order concerning the use of the renormalized single particle polarizabilities in the above formulae. In simulations, it is recognized that the application of the CDM equations needs of some \emph{a priori} prescription
for the value of the renormalized single-particle
polarizability in order to obtain physically acceptable values
for the scattering cross-section \cite{Sentenac3}. As a first approximation, one can use polarizabilities which are only renormalized by  in-free-space radiative corrections --see eg.\cite{Sentenac}. \emph{A priori} additional renormalizations were carried out by Draine and Goodman in \cite{Draine2} and Chaumet \emph{et al.} in \cite{Sentenac3}. Nonetheless, it must be clear by now that because the $\varphi$-factors depend on the renormalized value of $\tilde{\alpha}$ and $\tilde{\alpha}$ receives radiative corrections from the $\varphi$-factors, the problem is one of self-consistency. In physical terms, this reflects the double and complementary role that actual dipoles play in a complex medium. On the one hand, they polarize the vacuum. On the other hand, they renormalize their own polarizabilities.

\section{Conclusions}
We have computed the propagator of the polarization field in a statistically homogeneous and isotropic medium made of indistinguishable isotropic point dipoles. In Fourier-space, a relation of proportionality with Dyson's propagator has been found in terms of local field factors --Eqs.(\ref{LDOSIper},\ref{LDOSIparal},\ref{decomp}). The stochastic kernel of its Lippmann-Schwinger equation is that of Eq.(\ref{Xi}). The self-polarization propagator determines the LDOS of the total dipole emission, $\mathcal{N}^{emis.}$.  In contrast, Dyson's propagator determines only the LDOS of the power transferred radiatively and directly from a dipole emitter into the medium. The latter is equivalent to the LDOS of the coherent light which propagates through the medium from an external source, $\mathcal{N}^{light}$. Correspondingly, an emission vacuum and a light vacuum are postulated. The symmetry group of the latter is included in that of the former.\\
\indent An expression for the optical theorem in complex media has been found for classical dipoles --Eq.(\ref{laWshort}). Formulae for the renormalized values of the resonant frequency, the decay rate and the bare polarizability have been derived in Eqs.(\ref{Gamares}-\ref{alpha0reg}) for a two-level atom. They agree with previous QM approaches.\\
\indent In the \emph{stricto sensu} virtual cavity scenario, stimulated emission has been classified in Eqs.(\ref{Gammapcoh}-\ref{Gammalincoh}) attending to its coherent/extinguished nature. Further, the coherent power has been decomposed in Eq.(\ref{Wcohy}) into a \emph{direct} and an \emph{indirect} component. The \emph{direct} component carries the energy transferred radiatively and directly from the emitter into the medium, its spectrum is $\mathcal{N}^{light}$ and its field is Dyson's. The \emph{indirect} component is that of the field radiated by the induced dipoles which is \emph{in-phase} with the Dyson field. Only one local field factor enters the coherent spectrum. Correspondingly, the coherent dipole field is proportional to the Dyson field, being the constant of proportionality a local field factor, Eq.(\ref{Coho}).\\
\indent Using Schwinger's variational method, an expression for the vacuum energy density has been obtained as a sum of different physical contributions, Eqs.(\ref{Flamb},\ref{bulky},\ref{lacagada}). The vacuum energy density of an infinite-volume effective medium has been computed. When local field factors are included, additional terms proportional to the resonant frequency and the line-width arise --Eq.(\ref{LFFG})-- in addition to the usual Schwinger-bulk term.\\
\indent The dependence on the refractive index of the density of different radiative modes in an effective medium has been analyzed. We have found that $\mathcal{N}^{light}\propto n$, $\mathcal{N}^{Sch.}_{bulk}\propto n^{3}$, $\mathcal{N}^{Coh.}_{MG}\propto(n^{3}+2n)/3$, $\mathcal{N}^{emis.}_{MG}\propto(n^{5}+4n^{3}+4n)/9$, where the last two expressions apply to off-resonant Maxwell-Garnett dielectrics only since $\mathcal{N}^{Coh.}$ and $\mathcal{N}^{emis.}$ are in general model-dependent.\\
\indent It is left for future work to verify whether the conjecture of the equivalence between the Lamb-shift and the total variation of the vacuum energy  holds. Should the result be negative, a physical interpretation for the discrepancy and possible observational signatures must be investigated.\\
\indent Also, it would be interesting to extend the present formalism to more general dielectric configurations. Beyond dipole-aggregate models it would be necessary to develop a multipole expansion similar to that in \cite{Jaffe,Emig}. Multipole sources and propagators should be defined and a similar diagrammatic analysis to the one performed here carried out.
\acknowledgments
We thank S.Albaladejo, L.Froufe, J.J.Saenz and C.Herdeiro for
fruitful discussions and suggestions. This work has been supported by the Spanish integrated
project Consolider-NanoLight CSD2007-00046, the EU project
 NanoMagMa EU FP7-NMP-2007-SMALL-1 and the Scholarship Program 'Ciencias de la Naturaleza' of the Ramon Areces Foundation.
\appendix
\section{Stimulated emission from a classical dipole}\label{append}
\begin{figure}[h]
\includegraphics[height=5.7cm,width=8.9cm,clip]{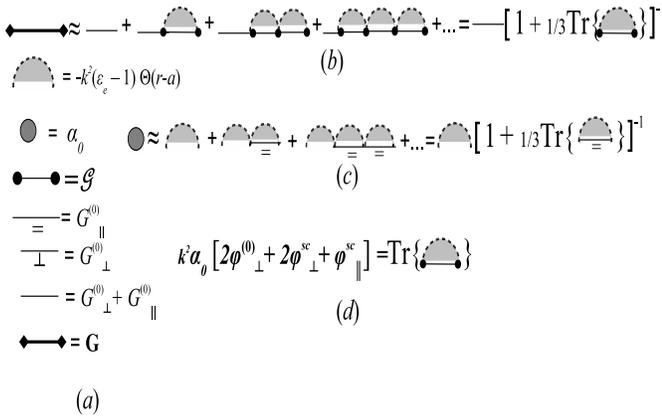}
\caption{($a$) Feynman's rules for the classical regularization scheme of Appendix \ref{append}. ($b$) Diagrammatic representation
of Eq.(\ref{b3}). ($c$) Diagrammatic representation of the
dressing up of $\chi_{e}$ leading to $\alpha_{0}$.
Approximation symbols denote that the field within the emitter is
taken uniform. ($d$) Diagrammatic representation of
 one of the self-polarization cycles which enter the series in $(b)$.}\label{fig21}
\end{figure}

We adopt a classical model in which the dipole emitter is a spherical nanoparticle of permittivity $\epsilon_{e}(\omega)$ and radius $a$, with $ka\ll1$ for the frequencies of interest.\\
\indent Let us consider that the emitter is embedded in a generic host medium and sits at position $\vec{r}$. It is stimulated by a stationary external field which oscillates in time with  frequency $\omega$, $\vec{E}_{0}(\vec{r})$. Formally, the averaged-power emitted reads \cite{Jackson},
\begin{equation}\label{lasimple}
W^{tot}_{\omega}=\frac{\omega}{2}\Im{\{\int\textrm{d}^{3}r\:\Theta(r-a)
\vec{\mathfrak{p}}(\vec{r})\cdot\vec{E}_{0}^{*}(\vec{r})\}},
\end{equation}
where $\vec{\mathfrak{p}}(\vec{r})$ is the polarization density induced on the emitter, which is proportional to $\vec{E}_{0}(\vec{r})$ in the linear-small-particle approximation and is affected by self-polarization effects. We introduce the self-polarization field through the insertion of appropriate Green's functions in the above expression,
\begin{eqnarray}
\vec{\mathfrak{p}}(\vec{r})&=&\int\textrm{d}^{3}r''
\Theta(r-a)\epsilon_{0}\chi^{\omega}_{e}\nonumber\\&\times&\int\textrm{d}^{3}r'
\bar{\mathrm{\mathbf{G}}}(\vec{r},\vec{r}';\omega)\cdot[\bar{G}^{(0)}]^{-1}(\vec{r}'-\vec{r}'';\omega)
\cdot\vec{E}_{0}(\vec{r}''),\nonumber
\end{eqnarray}
so that Eq.(\ref{lasimple}) reads
\begin{eqnarray}\label{lacomplex}
W^{tot}_{\omega}&=&
\frac{\omega}{2}\Im{}\Bigl\{\int\textrm{d}^{3}r\:\chi^{\omega}_{e}\:\Theta(r-a)\int\textrm{d}^{3}r'
\textrm{d}^{3}r''
\bar{\mathrm{\mathbf{G}}}(\vec{r},\vec{r}';\omega)\nonumber\\&\cdot&[\bar{G}^{(0)}]^{-1}(\vec{r}'-\vec{r}'';\omega)
\cdot\vec{E}_{0}(\vec{r}'')\cdot
\vec{E}^{*}_{0}(\vec{r})\Bigr\}.
\end{eqnarray}
We drop the script $\omega$ hereafter unless necessary.
In the above formulae $\chi^{\omega}_{e}=\epsilon_{e}(\omega)-1$ is the relative electrostatic susceptibility of the emitter --not to be
confused with the susceptibility of the host medium-- and
\begin{equation}\label{b3}
\bar{\mathrm{\mathbf{G}}}(\vec{r})\approx\bar{G}^{(0)}(\vec{r})
\sum_{m=0}^{\infty}\Bigl[-k^{2}\chi^{\omega}_{e}\int\Theta(v-a)\bar{\mathcal{G}}(v)\textrm{d}^{3}v\Bigr]^{m}
\end{equation}
is the propagator which takes account of the infinite number of self-polarization cycles which give rise to radiative corrections.  $\bar{\mathrm{\mathbf{G}}}(\vec{r},\vec{r}')$ propagates virtual photons from a point $\vec{r}'$ inside the emitter back to another point $\vec{r}$ also within the emitter.  All the equations above become simple in the small particle limit, $a\ll k^{-1}$, for the electric field
is nearly uniform within the emitter and so are the polarization density and the propagator
$\bar{\mathrm{\mathbf{G}}}(\vec{r},\vec{r}')$. The $n$-point irreducible diagrams which enter the computation of $\bar{\mathrm{\mathbf{G}}}$
can be approximated by the series of Fig.\ref{fig21}($b$) in which the two-point correlation functions $\Theta(r-a)$ appear consecutively as factors of a product. That way the corresponding integrals appear untangled and the series becomes  geometrical. Hence, Eq.(\ref{b3}). The underlying approximation reads,
\begin{eqnarray}
&\int&\textrm{d}^{3}r\:\Theta(r-a)\bar{\mathcal{G}}(r)\simeq\frac{4\pi}{3}a^{3}\bar{\mathcal{G}}(0)\nonumber\\
&=&\frac{4\pi}{3}a^{3}\Bigl[2\varphi_{\perp}^{(0)}+\varphi_{\parallel}^{(0)}+
2\varphi^{sc}_{\perp}+\varphi^{sc}_{\parallel}\Bigr]\frac{1}{3}\bar{\mathbb{I}}.
\end{eqnarray}
It was mentioned in Subsection \ref{2B} that  $2\varphi_{\perp}^{(0)}$ and $\varphi_{\parallel}^{(0)}$ are divergent. The divergence of $\varphi^{(0)}_{\parallel}$ is cured by the presence of the finite radius $a$. Since the limit   Lim$\{\int\textrm{d}^{3}r\:\Theta(r-a)\bar{G}_{stat.}^{(0)}(r;\omega)\}=\frac{1}{3k^{2}}\bar{\mathbb{I}}$ as $ka\rightarrow0$ is conditionally convergent, it is the Heaviside function of the integrand that yields the finite value $\frac{1}{3k^{2}}\bar{\mathbb{I}}$ \cite{Yagh}. Any other geometry would give a different numerical value. By equating that result with  $\frac{4\pi}{3}a^{3}\varphi^{(0)}_{\parallel}\frac{1}{3}\bar{\mathbb{I}}$ (leaving $\Re{\{2\varphi^{(0)}_{\perp}\}}$ still free)
we obtain $\varphi^{(0)}_{\parallel}=(\frac{4\pi}{3}a^{3}k^{2})^{-1}$. The net effect of this regularization procedure is to dress up the single particle
susceptibility in all the in-free-space electrostatic corrections. This procedure is depicted in Fig.\ref{fig21}($c$).
That way we can define
$\tilde{\chi}^{\omega}_{e}\equiv\frac{3}{\epsilon_{e}(\omega)+2}\chi^{\omega}_{e}$ and obtain the bare electrostatic polarizability $\alpha_{0}(\omega)\equiv 4\pi
a^{3}\frac{\epsilon_{e}(\omega)-1}{\epsilon_{e}(\omega)+2}$.
With the above
definitions we can rewrite Eq.(\ref{lacomplex}) in terms of
electrostatically renormalized operators,
\begin{eqnarray}\label{laleche2}
W^{tot}_{\omega}&=&
\frac{\omega\epsilon_{0}}{2}\Im{}\Bigl\{\int\textrm{d}^{3}r\:\tilde{\chi}^{\omega}_{e}\:\Theta(r-a)\int\textrm{d}^{3}r'
\textrm{d}^{3}r''
\bar{\tilde{\mathrm{\mathbf{G}}}}(\vec{r},\vec{r}')\nonumber\\&\cdot&[\bar{G}^{(0)}]^{-1}(\vec{r}'-\vec{r}'')
\cdot\vec{E}_{0}(\vec{r}'')\cdot
\vec{E}^{*}_{0}(\vec{r})\Bigr\},
\end{eqnarray}
where
\begin{equation}\label{b4}
\bar{\tilde{\mathrm{\mathbf{G}}}}(\vec{r},\vec{r}')\equiv\bar{G}^{(0)}(\vec{r}-\vec{r}')
\sum_{m=0}^{\infty}(\frac{-k^{2}\alpha_{0}}{3})^{m}[2\varphi^{(0)}_{\perp}+2\varphi^{sc}_{\perp}+\varphi^{sc}_{\parallel}]^{m}.\nonumber
\end{equation}
Inserting the above equation into Eq.(\ref{laleche2}) we obtain the expressions of Eqs.(\ref{la2},\ref{la3}) and the renormalized polarizability of Eq.(\ref{alpha1}).
\section{Renormalization of a two-level atom isotropic polarizability}\label{appendy}
The bare polarizability of a two-level atom  reads,
\begin{equation}
\bar{\alpha'}=\frac{2\omega_{0}}{\hbar\epsilon_{0}}\frac{\vec{\mu}\otimes\vec{\mu}}{\omega_{0}^{2}-\omega^{2}},
\end{equation}
where $\vec{\mu}\otimes\vec{\mu}$ is the tensor product of the dipole-dipole transition matrix elements. The EM interactions which give rise to the atomic bound state have been integrated out and parametrized by the resonant frequency, $\omega_{0}$. This way, only coupling to radiative modes is still missing.
In a spherically symmetric state, $\vec{\mu}\otimes\vec{\mu}=\mu^{2}/3\bar{\mathbb{I}}$, we can write $\bar{\alpha}'=\frac{1}{3}\frac{\alpha_{0}\omega_{0}^{2}}{\omega_{0}^{2}-\omega^{2}}\bar{\mathbb{I}}$ with $\alpha_{0}=\frac{2|\mu|^{2}}{\epsilon_{0}\hbar\omega_{0}}$. The self-polarization of the atom amounts to its coupling to radiative modes. Each self-polarization cycle carries a factor
\begin{eqnarray}
-k^{2}\textrm{Tr}\{\bar{\alpha}'\cdot\bar{\mathcal{G}}(\vec{r},\vec{r})\}&=&\frac{-k^{2}}{3}\frac{\alpha_{0}\omega_{0}^{2}}{\omega_{0}^{2}-\omega^{2}}
\nonumber\\&\times&[2i\Im{\{\varphi^{(0)}_{\perp}\}}+2\varphi^{sc}_{\perp}+\varphi^{sc}_{\parallel}].
\end{eqnarray}
The infinite sum of cycles is a geometrical series analogous to that of Fig.\ref{fig21}($b$) for classical dipoles. It amounts to $1/3$ times Eq.(\ref{alpha1}) with the above definitions.

\end{document}